\documentclass[fleqn,usenatbib]{mnras}

\usepackage{newtxtext,newtxmath}

\newcommand{\Ha}{$\rm{H} \alpha$}
\newcommand{\Hb}{$\rm{H} \beta$}

\newcommand{\HI}{\hbox{{\rm H}\kern 0.1em{\sc i}}}
\newcommand{\Lya}{\hbox{{\rm Ly}\kern 0.1em$\alpha$}}
\newcommand{\Lyb}{\hbox{{\rm Ly}\kern 0.1em$\beta$}}
\newcommand{\MgII}{\hbox{{\rm Mg}\kern 0.1em{\sc ii}}}
\newcommand{\SiII}{\hbox{{\rm Si}\kern 0.1em{\sc ii}}}
\newcommand{\SiIII}{\hbox{{\rm Si}\kern 0.1em{\sc iii}}}
\newcommand{\SiIV}{\hbox{{\rm Si}\kern 0.1em{\sc iv}}}
\newcommand{\CII}{\hbox{{\rm C}\kern 0.1em{\sc ii}}}
\newcommand{\CIII}{\hbox{{\rm C}\kern 0.1em{\sc iii}}}
\newcommand{\CIV}{\hbox{{\rm C}\kern 0.1em{\sc iv}}}
\newcommand{\NII}{\hbox{{\rm N}\kern 0.1em{\sc ii}}}
\newcommand{\NIII}{\hbox{{\rm N}\kern 0.1em{\sc iii}}}
\newcommand{\NV}{\hbox{{\rm N}\kern 0.1em{\sc v}}}
\newcommand{\OVI}{\hbox{{\rm O}\kern 0.1em{\sc vi}}}
\newcommand{\OII}{\hbox{[{\rm O}\kern 0.1em{\sc ii}]}}
\newcommand{\OIII}{\hbox{[{\rm O}\kern 0.1em{\sc iii}]}}
\newcommand{\kms}{\hbox{~km~s$^{-1}$}}

\newcommand{\colden}{\ensuremath{\log (N(\HI)/{\rm cm}^{-2})}}
\newcommand{\fewcorot}{\hbox{$f_{\rm EWcorot}$}}
\newcommand{\slope}{\hbox{${d{(\fewcorot)}}/{d{(\rm eV)}}$}}

\newcommand{\dfcorot}{\hbox{$\Delta \fewcorot (\rm{\MgII} - \rm {\OVI})$}}

\usepackage[T1]{fontenc}
\usepackage{booktabs} 
\DeclareRobustCommand{\VAN}[3]{#2}
\let\VANthebibliography\thebibliography
\def\thebibliography{\DeclareRobustCommand{\VAN}[3]{##3}\VANthebibliography}

\usepackage{graphicx}	
\usepackage{amsmath}	
\usepackage{multirow}
\usepackage{setspace}
\usepackage{bigdelim}
\usepackage{bigstrut}
\usepackage{floatflt}	
\usepackage{wrapfig}
\usepackage{subfig}
\usepackage{multicol}
\usepackage[utf8]{inputenc}
\usepackage[flushleft]{threeparttable}
\usepackage{xltabular}
\usepackage{caption}
\usepackage{lscape}

\defcitealias{GF1}{GasFlows-I}

\title[Galaxy--Multiphase CGM kinematics]{Signatures of Gas Flows–II: Connecting the kinematics of the multiphase circumgalactic medium to galaxy rotation}

\author[Nateghi et al.]
{Hasti Nateghi,$^{1,2}$\thanks{E-mail: hnateghi@swin.edu.au}
Glenn G. Kacprzak$^{1,2}$,
Nikole M. Nielsen$^{1,2,3}$, 
Sameer$^{4,5}$, 
Michael T. Murphy$^{1}$,
\newauthor Christopher W. Churchill$^{6}$, 
Jane C. Charlton$^{5}$\\
$^{1}$Centre for Astrophysics and Supercomputing, Swinburne University of Technology, Hawthorn, Victoria 3122, Australia\\
$^{2}$ARC Centre of Excellence for All Sky Astrophysics in 3  Dimensions (ASTRO 3D), Australia\\
$^{3}$Homer L. Dodge Department of Physics and Astronomy, The University of Oklahoma, 440 W. Brooks St., Norman, OK 73019, USA\\
$^{4}$Department of Physics and Astronomy, The University of Notre Dame, Notre Dame, IN 46544, USA\\
$^{5}$Department of Astronomy and Astrophysics, The Pennsylvania State University, State College, PA 16801, USA\\
$^{6}$Department of Astronomy, New Mexico State University, Las Cruces, NM 88003, USA
}

\date{Accepted 2024 September 9. Received 2024 September 9; in original form 2023 November 3}

\pubyear{2024}

\begin{document}
\label{firstpage}
\pagerange{\pageref{firstpage}--\pageref{lastpage}}
\maketitle

\begin{abstract}

The multiphase CGM hosts critical processes that affect galaxy evolution such as accretion and outflows.  We searched for evidence of these phenomena by using the EW co-rotation fraction (\fewcorot) to study the kinematic connection between the multiphase CGM and host galaxy rotation. We examined CGM absorption from HST/COS (including, but not limited to, {\SiII}, {\CII}, {\SiIII}, {\CIII}, and {\OVI}) within $21\leq D\leq~276$~kpc of 27 galaxies. We find the median {\fewcorot} for all ions is consistent within errors and the {\fewcorot} increases with increasing N$({\HI})$. The {\fewcorot} of lower ionization gas decreases with increasing $D/R_{\rm vir}$ while {\OVI} and {\HI} are consistent with being flat. The {\fewcorot} varies minimally as a function of azimuthal angle and is similar for all ions at a fixed azimuthal angle. The larger number of {\OVI} detections enabled us to investigate where the majority of co-rotating gas is found. Highly co-rotating {\OVI} primarily resides along the galaxies' major axis. 
Looking at the {\fewcorot} as a function of ionization potential ({\slope}), we find a stronger co-rotation signature for lower-ionization gas. There are suggestions of a connection between the CGM metallicity and major axis co-rotation where low-ionization gas with higher {\fewcorot} exhibits lower metallicity and may trace large-scale filamentary inflows. Higher ionization gas with higher {\fewcorot} exhibits higher metallicity and may instead trace co-planar recycled gas accretion. Our results stress the importance of comparing absorption originating from a range of ionization phases to differentiate between various gas flow scenarios. 
\end{abstract}




\section{Introduction}

The CGM is a multiphase gaseous halo surrounding galaxies that hosts the baryon cycle which is primarily driven by accretion and outflows \citep{Tumlinson11, Tumlison17}. It holds significant reservoirs of neutral hydrogen and metals beyond the interstellar medium (ISM) of galaxies, extending outwards to the virial radius  \citep[e.g.,][]{chen2010may, stocke2013, Tumlison17}. Many studies have investigated the process in which gas travels through the CGM and approaches the disk to join the ISM to feed galaxy star formation. Hot and cold mode accretion are the best candidates for this process \citep[e.g.,][]{keres2005,keres2009, Dekel_Birnboim2006,FaucherGiguere&keres2011,vandeVoort2011,AnglsAlczar2017, strawn2021,Hafen2022, Afruni2023}. Cold mode accretion dominates in lower mass ($\log (M_{\rm h}/M_{\odot})\lesssim 12$) galaxies, moving through the CGM as dense anisotropic filaments \citep{keres2005, Dekel2009, Faucher-Gigu2011a,stewart2017_revi}, whereas the hot mode is more dominant in massive halos ($\log (M_{\rm h}/M_{\odot}) \gtrsim 12$) where the gas virializes to the temperature of the halo and infalls isotropically \citep{Katz2003,vandeVoort2011, vandeoortSchay2012,Feilding2017, Hafen2022}.


Cosmological simulations predict that the accretion of cold filaments occurs within an extended co-rotating disk aligned with the galaxy plane \citep{stewart2011a, stewart2011b, stewart2013, Danovich2015,Nelson2016,stewart2017, suresh2019, peroux2020b}. Connecting the kinematics of the CGM to the host galaxy rotation could then reveal signatures of gas flows. Utilising absorption lines present in the spectra of background quasars is still the best observational method to study the CGM and its kinematics. This approach enables us to unveil the multiphase gas in the CGM with different temperatures and densities, and to study their relative kinematics \citep[e.g.,][]{Morris93, Tripp_1998, Glenn2008,steidel2010, Rudie12, werk2014, Peroux2019, Nikki2020_cosmicnoon, Hasti2021}.
Indeed, the Cosmic Origins Spectrograph (COS) on {\it Hubble Space Telescope} ({\it HST}) enables us to detect numerous metal species, including \CII, \CIII, \CIV, \SiII, \SiIII, \SiIV, \NV, and \OVI, tracing low-, intermediate-, and high ionization phases of the CGM. These absorption lines allow for an examination of the physical conditions and kinematics of the gas surrounding galaxies, especially in the low redshift Universe where galaxies are easy to identify. 

The observational evidence for gas accretion is best found in studies of low ionization {\MgII} absorption along the major axes of galaxies that frequently reveal Doppler-shifted absorption with the same direction as the galaxy rotation \citep{Steidel2002, Glenn2010a, Bouche13, bouche2016,Diamond-Stanic2016,  Ho17, hadi18,MartinCrystal2019, Lopez2020}. This indicates that the low-ionization CGM most likely co-rotates with the galaxy disk even at substantial distances from the galaxy ($D\sim100$~kpc), consistent with simulation predictions \citep[e.g.,][]{stewart2013}. 
Using $\Lambda$CDM simulations \citet{Glenn2010a} showed that the {\MgII} absorbing gas mostly originates from the filaments and tidal streams with inward velocities, falling into the host galaxies. Furthermore, \cite{Ho2019} used EAGLE simulations to indicate anisotropic accretion onto the galaxies where the gas is located within 10 deg and 60 kpc of the galaxy disk.

Given the multiphase nature of the CGM, the connection between the galaxies' rotational spin and the warm/hot halo is also important but less explored. From observations \citep[e.g.,][]{muzahid15,werk2016, Nikki2017,Hasti2021} and simulations \citep[e.g.,][]{churchill2015,Ford2014,Ford2016}, we often expect different kinematics and absorption profiles for higher-ionization absorbers. Previous work has found that there was no strong correlation between the kinematic spread of gas and galaxy properties for the high ionization phase traced by {\OVI} \citep{Glenn2015decmorpho, Nikki2017,Glenn2019,MasonNg2019}. Nevertheless, \citet{Glenn2019} found that there might be some observational connection between the kinematics of {\OVI} absorption and galaxy rotation. They further used cosmological simulations to probe the kinematics of the gas which showed evidence of inflows along the major axis. However, they concluded that the observational kinematic signatures of {\OVI} in accretion could be obscured by the broad distribution of the {\OVI} velocities across the halo.

Studying {\HI} kinematics \citep{Barcon1995, Cote2005, French2020, Klimenko23, GF1} may explain the disparity in the aforementioned studies and bridge the gap between the low- and high-ionization halos. Since the low-ionization phase of the CGM ({\MgII}) is mainly associated with strong {\HI} absorption, while the high-ionization phase ({\OVI}) is found for the entire {\HI} column density range, the difference in co-rotation for {\MgII} and {\OVI} would likely be reflected in weak and strong {\HI} systems. Additionally, the absorption systems in the {\MgII} and {\OVI} studies were categorised solely as either co-rotating or not based on whether the bulk of the absorption was consistent with the galaxy's rotation direction, without providing a quantification of the amount of gas in each system that could potentially co-rotate with the galaxy. As the {\OVI} results suggest, there are likely many gas flows present along a given line-of-sight that could obscure accretion kinematic signatures.

In the first paper of this series, \citet[][hereafter \citetalias{GF1}]{GF1}, we addressed these shortcomings by studying a sample of 70 galaxy--quasar pairs that exhibit {\HI} absorption with a large range of {\HI} column densities ($12<{\colden}<20$). We quantified the amount of gas that is consistent with galaxy rotation using the equivalent width co-rotation fraction, {\fewcorot}. Our findings indicated a correlation between the co-rotation fraction of {\HI} and its column density, which is consistent with the {\MgII} (high $N({\HI})$ only) and {\OVI} (range of $N({\HI})$) results in the literature. We also found that the highest co-rotation fractions were located within the virial radius ($R_{\rm vir}$) of galaxies, where metal lines are predominantly found, although high co-rotation fractions were also found outside $R_{\rm vir}$ but only along the major axis (consistent with IGM accreting filament scenarios). Despite these results and the literature, it is still debated how the relative galaxy--CGM kinematics differs with ionization.

In this study, we aim to constrain the relative kinematics between the CGM and galaxies by investigating variations in CGM absorbing gas co-rotation across different ionization states. We use a subset of the \citetalias{GF1} sample which has a range of low-, intermediate-, and high-ionization metal lines to characterise how the kinematics depend on ionization. The paper is organised as follows: 
In Section~\ref{sec:method} we elaborate on the sample and our analysis.  In Section~\ref{sec:results} we present and explore the results of how {\fewcorot} varies as a function of the {\HI} column density, impact parameter normalised to the virial radius, azimuthal angle, and ionization potential of different CGM species. We present our concluding remarks in Section~\ref{sec:conslusion}.
Throughout this paper, we adopt an ${\rm H}_{\rm 0}=70$~\kms~Mpc$^{-1}$, $\Omega_{\rm M}=0.3$, $\Omega_{\Lambda}=0.7$ cosmology.

\section{Sample \& Method}
\label{sec:method}
A sample of 27 galaxy--CGM metal absorption pairs ($0.09<z<0.5$, $z_{\rm median}=0.23\pm0.10$) is used to examine their kinematic connection in different ionization states.  This is a subsample of the larger dataset consisting of 70 {\HI} pairs examined in \citetalias{GF1}. Each of the 27 galaxies studied here contains at least one detected metal-line detected in the spectrum of a single background quasar ($D=21-276$~kpc; only one sightline per galaxy) observed with {\it HST}/COS and Keck/HIRES or VLT/UVES.  Their {\HI} column densities are shown as a function of impact parameter in Fig.~\ref{fig:sample} along with the distribution of halo masses. The galaxies have a halo mass range of $\log (M_{\rm h}/M_{\odot})=10.5-12.3$ and their rotation curves were obtained with the Echelle Spectrograph and Imager \citep[ESI,][]{Sheinis2002} on Keck~II as part of the analysis in \citetalias{GF1}. Our sample contains both galaxy-selected \citep{pointon19} and absorption-selected \citep{Tripp2008} absorber--galaxy pairs.  

Interactions in group environments or major mergers can complicate the kinematic connection between the galaxy and CGM. For example, there could be perturbations on the galaxy rotation curves or gas distributions due to interactions or major mergers. Because of this, our analysis primarily centres on isolated galaxies to minimise the influence of these external factors \citepalias{GF1}. Our galaxies were selected from \citep{pointon19}, who state that there are no major companions within 100~kpc or with velocity separations less than 500~km~s$^{-1}$ of their galaxies. The galaxies may still have nearby minor companions, which likely do not affect the kinematics of the larger galaxy.

Details regarding the metal lines observations and the galaxy sample can be found in Tables~\ref{galaxy-abs} and \ref{EW-detections}.  We elaborate on the specifics of the data and analysis for this sample in the subsequent subsections.

\subsection{Quasar Observations}
\label{QSO obs}

\subsubsection{Spectroscopy}

{\it HST}/COS was used to obtain the UV spectra of background quasar in each field. The G130M and/or G160M spectra have a moderate resolving power of $R\sim20,000$,  giving a full width at half maximum of $\sim18$~{\kms} and wavelength coverage of $1410-1780$~{\AA}. Further details regarding the QSOs spectroscopy and reductions are provided in Sameer et al. (in prep). In brief,  the raw spectra were processed using the CALCOS V3.2.1 pipeline and then aligned and combined following the method described in \citet{Wakker2015}. The spectra are heliocentric corrected and have vacuum wavelength. 
When available, for 10/27 systems,  we used Keck/HIRES or VLT/UVES quasar spectra to complement our {\it HST}/COS spectra by including {\MgII} absorption. The HIRES spectra were reduced using the Maunakea Echelle Extraction package or IRAF. The UVES spectra were reduced using the European Southern Observatory pipeline \citep{Dekker2000UVES} and the UVES Post-Pipeline Echelle Reduction (UVES POPLER) software \citep{Murphy2019}. All optical spectra have vacuum wavelengths and are heliocentric corrected.

\begin{figure}
    \centering
    \includegraphics[width=\columnwidth]{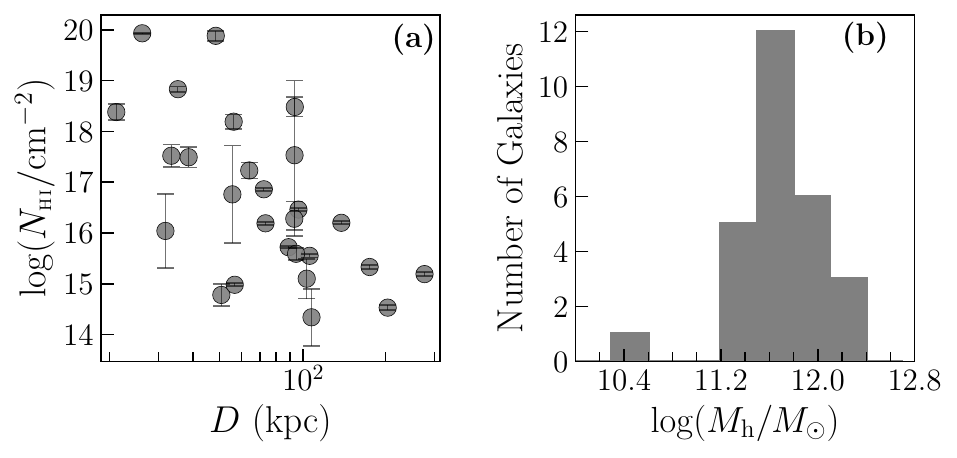}\hfill 
    \caption{ (a) Distribution of {\HI} absorption column densities in our sample as a function of projected distance from galaxies. (b) The halo mass distribution of the galaxies in our sample.
    }
    \label{fig:sample}
\end{figure}

Neutral hydrogen column densities are adopted from either \citet{sameer2024}, \cite{pointon19}, \cite{Tripp2008} or \citetalias{GF1} (see Table A1 in \citetalias{GF1} for the references of the column densities) and range between $\log (N({\HI})/{\rm cm}^{-2})=13.7 - 20$. Table~\ref{galaxy-abs} contains the {\HI} column densities, which are also shown in Fig.~\ref{fig:sample}.

\subsubsection{Multiphase CGM Detections}

We selected 27 galaxies from \citetalias{GF1}, each hosting one or more of the following absorption lines:
{\MgII~$\lambda 2796$, \CII~$\lambda 1334$, \CIII~$\lambda 977$, \CIV~$\lambda 1548$, \SiII~$\lambda 1260$, \SiIII~$\lambda 1206$, \SiIV~$\lambda 1393$, \NII~$\lambda 1083$, \NIII~$\lambda 989$, \NV~$\lambda 1238$, and \OVI~$\lambda 1031$}. It is important to note that not all ions are present in each system and Table \ref{EW-detections} provides the detection information for individual galaxies.  Among the detected transitions, {\SiII} (16.3~eV, the ionization energy required to remove an electron from the neutral atom to create the ion), {\CII} (24.4~eV), {\SiIII} (33.5~eV), {\CIII} (47.9~eV), and {\OVI} (138.1~eV) had the highest detection rates, enabling us to perform statistical comparisons. Out of 27 systems, {\OVI} was present in 25, {\SiIII} in 21, {\CIII} in 19, {\CII} in 18 and {\SiII} in 15.  We also have 23 galaxies that host 3 or more metal-lines where we can further examine how the kinematics varies across different ions.  

In Sections~\ref{3.1}, \ref{3.2}, and \ref{3.3}, we analyse how the kinematics of CGM, traced by these ions, relate to {\HI} strength, projected distances from the host galaxies, and orientation. In Section~\ref{3.4}, we investigate how galaxy--CGM relative kinematics vary with ionization within individual systems, considering all of the aforementioned detected transitions. We explore whether these variations depend on galaxy orientation or CGM metallicity.

\subsection{Galaxy Observations and Properties}
\label{galaxy obs}

\subsubsection{Spectroscopy}
The galaxy redshift zeropoints and their rotation curves are necessary to conduct this study of the galaxy--multiphase CGM relative kinematics. 
We have observed the 27 galaxies discussed in this work using the Keck/ESI with a slit of $1''$ wide and $20''$ long. For each galaxy, the slit was aligned with the projected optical major axis of the galaxy to obtain the full range of rotation velocities as further explained in \citetalias{GF1}. We used the standard echelle package in IRAF to combine the exposures and extract the spectra. The obtained echellette spectra have a wavelength coverage of 4000 -- 11000~{\AA} and are vacuum wavelength and heliocentric velocity corrected. Our ESI spectra have a spectral resolution of $R\sim4600$ with a sampling rate of 22~{\kms} pixel$^{-1}$ (${\rm FWHM}\sim65$~{\kms}). 

We extracted the rotation curves of the galaxies using the same approach adopted by \citet{Glenn2010a} 
which is similar to \cite{vogt1996} and \cite{steidel2010}. In this method, we employed a three-pixel-wide aperture size and shifted it at one-pixel intervals along the spatial direction to extract a sequence of spectra along the major axis of each galaxy. Gaussian fits were conducted on the galaxy emission lines, primarily {\Ha} and {\Hb}, to obtain wavelength and line of sight velocity centroids. These centroids were then utilised to determine the systemic redshifts and rotation curves of the galaxies. The redshifts of the galaxies are presented in Table~\ref{galaxy-abs}.

\subsubsection{Imaging and Morphology}
To establish a connection between the host galaxies and CGM absorption, we require information about the galaxy morphologies including inclination angles ($i$) and the orientation of the quasar sightline with respect to the galaxies' disk (azimuthal angles, $\Phi$). The $\Phi=0$ and $\Phi=90$ indicate the sightlines along the galaxies' major and minor axes respectively.  
To measure these properties, we used images of each field having either ACS, WFC3, or WFPC2 on {\it HST} with the F702W, F814W, or F625W filters to model the galaxy morphologies/geometries (see Table B1 in \citetalias{GF1} for imaging details). The inclination and azimuthal angles for each galaxy were modelled using the same method used in \citet{Glenn2011augustGIM2D} and \citet{Glenn2015decmorpho}. In brief, this approach determined the orientation and morphological characteristics of galaxies by employing two-component models that include both a disk and a bulge. These models were implemented using the GIM2D software \citep{Simard2002}. The galaxies' inclination and azimuthal angles are presented in Tables~\ref{galaxy-abs}.

\subsubsection{Mass and Virial Radius}
We adopt the halo mass and virial radius of the galaxies from \citetalias{GF1} where we convert the computed galaxy stellar masses ($M_{\ast}$) to halo masses ($M_{\rm h}$) using the stellar-to-halo mass relation from \citet{Girelli2020}. In summary, to compute the $M_{\ast}$ for galaxies in this paper, we obtained Kron magnitudes in the $g$ and $r$ filters from  Pan-STARRS \citep{chambers16}. We applied $K$-corrections following the procedures outlined in \citet{Nikki13a_magiicat1} to derive the rest-frame absolute magnitudes in the $g$ and $r$ bands. For the 15 galaxies without observed $g-r$ colours, we made the assumption of an Sbc-type spectrum. The requirement that all of our galaxies have rotation curves derived from nebular emission lines means that they are emission line-selected and are therefore star-forming. This is consistent with the Sb-Sc type morphologies seen in the {\it HST} imaging of these galaxies, however we are unable to accurately classify them from these images alone. This choice also aligns with the typical coloration of galaxies associated with CGM absorption, as supported by previous studies \citep{Steidel94, zibetti2007, Nikki13a_magiicat1, Glenn2015decmorpho}. 

We calculated the stellar masses for all galaxies by employing the rest-frame $g-r$ colour and the $g$-band mass-to-light ratio ($M/L$) relation as described in \citet{Bell2003}. We did not apply corrections for dust extinction since they are within the range of uncertainties associated with our method. We find good agreement with our stellar masses for 11 galaxies that overlap with the COS-Halos sample \citep{werk2013}, yielding a mean difference of 0.065~dex between the two samples.

The galaxies' virial radii, $R_{\rm vir}$, were determined using the methodology described in \citet{Bryan1998}. The halo masses and impact parameters normalised to the galaxies' virial radii ($D/R_{\rm vir}$) can be found in Table~\ref{galaxy-abs}. The distribution of halo masses are also shown in Fig~\ref{fig:sample}.

\subsection{Co-rotation Fractions}
\label{corot-fraction}


We employ the method developed in \citetalias{GF1} to measure the absorption equivalent width (EW) co-rotation fraction. This method allows us to quantify the fraction of the absorption EW caused by the CGM gas that has velocities consistent with the rotation direction of the host galaxies. The advantage of this approach is that it relies more on the data itself rather than user-dependent absorption decomposition.

To first measure the total EW of each metal absorption line, we used the profile fits of the observed data, which were adopted from \citet{sameer2024}, in order to exclude any blends and obtain accurate profile boundaries. To fit the absorption profiles, they implement a multiphase Bayesian ionization modeling technique that operates on a cloud-by-cloud basis. This approach has been utilised to extract the physical properties of the absorption systems and shown to yield reliable measurements \citep{sameer2021, sameer2022, sameer2024}.

For each galaxy, we then define a velocity window for its absorption lines, which includes all of the gas starting from the galaxy's systemic velocity to the edge of the absorption that resides in the direction of the rotation of the galaxy. We calculate the equivalent width in that window, which represents the absorption that aligns with the galaxy's spin direction along the quasar line of sight. We then divide that number by the total equivalent width of the absorption line and this quantity is referred to as the EW co-rotation fraction, $f_{\rm EWcorot}$. A value of 1 indicates that all of the gas is consistent with a co-rotation scenario, while a value of 0 suggests that none of the gas is consistent with such a scenario. Figure~2 in \citetalias{GF1} describes this method. Typical errors on the co-rotation fraction were calculated by bootstrapping the errors associated with galaxy redshift and the absorption profile and range from $0.001-0.006$.

\section{Results}
\label{sec:results}
We examine the kinematic relationship between galaxies and their surrounding multiphase CGM to quantify the fraction of gas that is consistent with possible co-rotation and/or accretion. In the following subsections, we explore the ions' co-rotation fractions, $f_{\rm EWcorot}$,  
as a function of neutral hydrogen column density, impact parameter normalised to the virial radius, and azimuthal angle. We then examine how the co-rotation fraction of the CGM varies as a function of ionization and metallicity for galaxies that have at least three different CGM metal ions.

\subsection{General statistics of co-rotation fractions}

\begin{figure}
    \centering
    \includegraphics[width=\columnwidth]{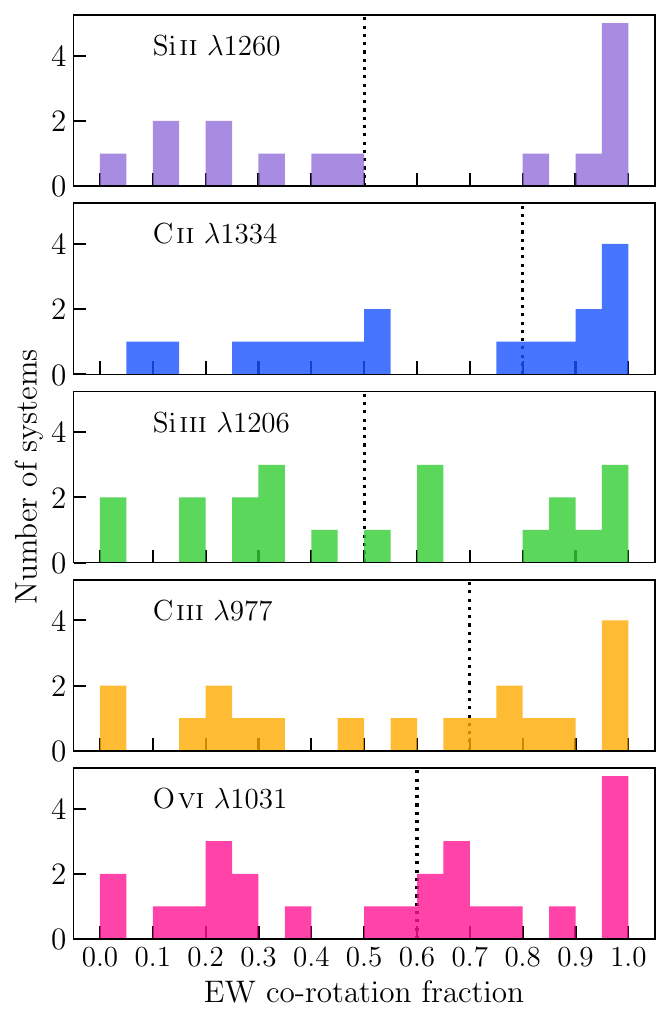}\hfill 
    \caption{ The co-rotation fraction distribution for our sample for a range of ions from the data listed in Table~\ref{EW corot}. Note that not all ions have the same number of systems (see text). The median is shown as a dotted black line. When the EW co-rotation fraction is greater than 0.5, we would consider that the bulk of the absorption is consistent with the rotation direction of the galaxy. 
    }
    \label{fig:ions-hist}
\end{figure}

Observations of {\MgII} absorption found that the bulk of the gas is consistent with galaxy rotation \citep[e.g.,][]{Steidel2002, Glenn2010a, Ho17, Zabl19_megaflow2},  while observations of {\OVI} show that only up to 50\% of the gas can be co-rotating with the galaxy \citep{Glenn2019}.  In Figure~\ref{fig:ions-hist}, we present the distribution of EW co-rotation fractions for a sample of ions. We find that there is a wide range of values ranging from 0 to 1 (see Table~\ref{EW corot}). We find that there is a skew toward unity, especially for {\SiII} and {\CII} and it seems to become more uniform for the higher ionization species. We further compute the median EW co-rotation fractions and the standard deviations for all of the systems, for a given ion, and find a value for {\SiII} of $0.5\pm0.4$, {\CII} of $0.8\pm0.3$, {\SiIII} of $0.5\pm0.3$, {\CIII} of $0.7\pm0.3$ and {\OVI} of $0.6\pm 0.3$.  Note that while there is some variation in the median values, they are still consistent. However, not all ions are covered in each system so some system-to-system variations are expected. The fact that our EW co-rotation fractions are consistent between ions is also supported by previous observations showing that {\OVI} exhibits similar, yet not identical, kinematics to {\HI} and low-ionization absorption \citep{Tripp2008,werk2014}.
\citet{MartinCrystal2019} also provide an example where the strongest component of {\OVI} has co-rotation consistent with {\MgII} absorption. We discuss how the co-rotation fraction of different ions varies for individual systems later in the Paper. 

We also highlight that {\fewcorot} values represent the statistical average of the fraction of the equivalent width that is consistent with being co-rotating with the galaxy for the population of galaxy-absorber pairs. For example, an average value $f_{\rm EWcorot}=0.5$ implies exactly half of the sample has most of their gas consistent with co-rotation and half inconsistent with co-rotation. This 50\% should not be interpreted as random, but as the fraction of the sample that exhibits co-rotation.  If we compare to the classical method \citep[i.e., identifies which systems have the bulk of their absorption to the correct side of the galaxy rotation;][]{Steidel2002,Glenn2010a,Ho17}  we find 7/15 (45\%) for {\SiII}, 11/18 (61\%) for {\CII}, 11/21 (52\%) for {\SiIII}, 11/19 (58\%) for {\CIII} and 15/25 (60\%) for {\OVI}, which is consistent with the average EW co-rotation fractions reported above. Thus, our overall results are comparable to other studies using different methods to quantify kinematic coherency between galaxies and their absorption.   

\subsection{Co-rotation of CGM phases versus \boldmath{\HI} column density}
\label{3.1}
\citetalias{GF1} investigated the {\fewcorot} of {\HI} while examining its relationship with {\HI} column density. They found a correlation between the {\HI} column density and its co-rotation fraction where the co-rotation increased with {\HI} column density.  This is consistent with simulations suggesting that Lyman Limit Systems (LLSs) could effectively trace CGM gas dynamics, including processes such as accretion and outflows \citep{vandeVoort2012, FaucherGiguere&keres2011, FaucherGiguere2015, Hafen2017}.

\begin{figure}
    \centering
    \includegraphics[width=\columnwidth]{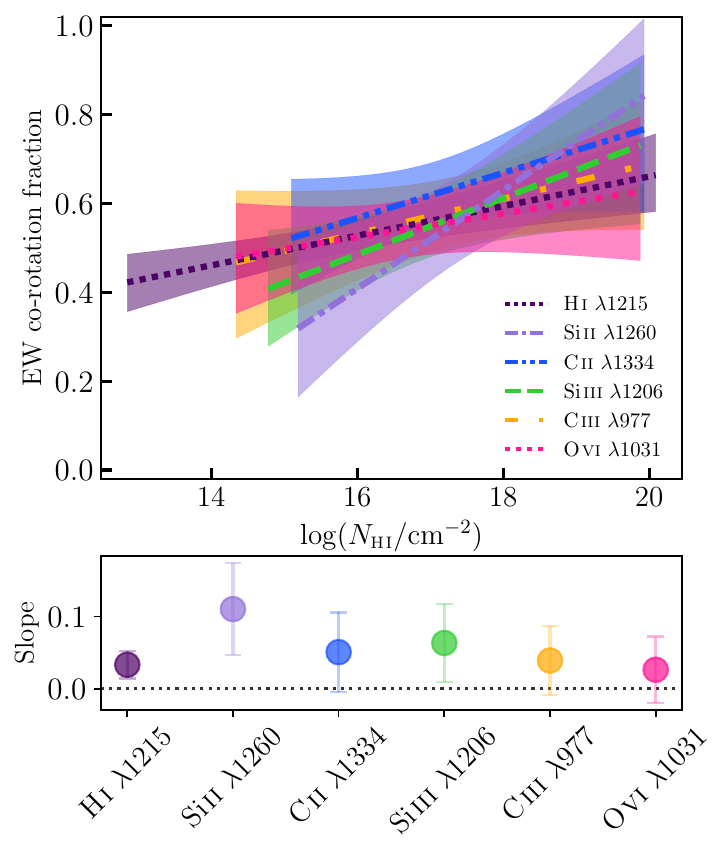}\hfill 
    \caption{(top) The polynomial fit to equivalent width co-rotation fraction ($f_{\rm EWcorot}$) of {\HI} and metal absorption lines as a function of {\HI} column density. The ions are colour-coded, ranging from cool to warm colours corresponding to increasing ionization potentials. The shaded regions represent the $1\sigma$ bootstrap uncertainties on the fits. (bottom) The slope of the fit to each ion and $1\sigma$ bootstrap errors. The colour of the points corresponds to the respective ion and fits in the top panel. Overall within the errors, the slopes appear to be consistent for all ions. }
    \label{fig:NHIcorot}
\end{figure}

Here we examine how the co-rotation fraction of different CGM ions is dependent on the {\HI} column density. The top panel of Figure~\ref{fig:NHIcorot} shows the first-order polynomial fit to $f_{\rm EWcorot}$ of each ion in relation to the {\HI} column density ($\colden$). We tested higher order polynomial fits, but the first-order fit best represents the data. The $1\sigma$ error of the fits, represented by the shaded area, is determined by bootstrapping the fit and subsequently calculating the average and standard deviation across 10,000 realisations. Errors on all values are included in the bootstrapping here and in subsequent analysis. The absorbers are colour-coded according to their ionization potential. Cool colours represent ions from the low-ionization phase of the CGM, while warmer colours represent ions from the higher phases. 
The dark purple shows the $f_{\rm EWcorot}$ for the 70 {\HI} absorbers from \citetalias{GF1}, the light purple shows the fit to the 15 {\SiII} absorbers, the blue shows the fit to the 18 {\CII} absorbers, the green shows the fit to the 21 {\SiIII} absorbers, the yellow shows the fit to the 19 {\CIII} absorbers and the pink shows the fit to the 25 {\OVI} absorbers. This figure shows the statistical average of the fraction of the equivalent width that is consistent with being co-rotating with the galaxy for the population of galaxy-absorber pairs.

We find that the co-rotation fractions of all ions are consistent within the errors with {\HI} co-rotation fraction, which increases with higher {\HI} column densities. The {\fewcorot} for all the ions goes from $\sim0.4$ to $\sim0.65$, for low to high column densities, respectively. 
To examine whether the {\fewcorot} versus $\rm N({\HI})$ varies with ionization, we explore the slopes of the fits to the $f_{\rm EWcorot}$ for each ion. In the bottom panel of Figure~\ref{fig:NHIcorot}, the slope of the fit for each ion along with the corresponding $1\sigma$ bootstrap errors is presented.  Overall within the errors, the slopes appear to be consistent. The data may hint that lower-ionization absorbers have a steeper slope, while higher-ionization absorbers and {\HI} have a more shallow slope, however we need more data to robustly determine this.

\begin{figure}
    \centering
    \includegraphics[width=\columnwidth]{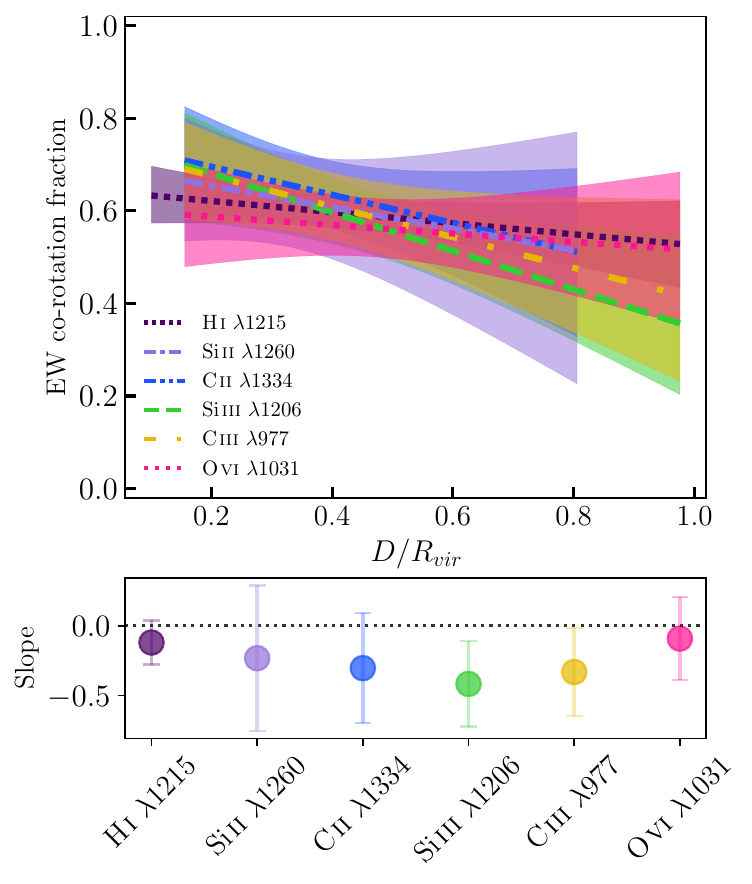}\hfill 
    \caption{(top) The polynomial fits to equivalent width co-rotation fraction (\fewcorot) as a function of $D/R_{\rm vir}$. The ions are colour-coded from cold to warm colours with increasing ionization potentials. The shaded regions represent the $1\sigma$ bootstrap uncertainties on the fits. (bottom) The slope of the fit to each ion and $1\sigma$ bootstrap errors. The colour of the points corresponds to the respective ion and fits in the top panel.}
    \label{fig:DRvircorot}
\end{figure}
\subsection{Co-rotation of CGM phases versus impact parameter}
\label{3.2}
Although we expect an anti-correlation between the CGM absorption strength and impact parameter, $D$ \citep[e.g.,][]{Tumlison17}, it is not clear how the co-rotation of metal absorbers behaves as a function of the galaxy--QSO projected separation.  \citetalias{GF1} investigated the {\fewcorot} of the {\HI} column density while examining its relationship with $D/R_{\rm vir}$. They found a flat distribution of {\fewcorot} for higher column density systems (that tend to reside within $R_{\rm vir}$) as a function of $D/R_{\rm vir}$. For low column density systems (that tend to reside beyond $R_{\rm vir}$), {\fewcorot} decreases with increasing $D/R_{\rm vir}$.

Figure~\ref{fig:DRvircorot}~(top) presents {\fewcorot} as a function of impact parameter normalised to the virial radius ($D/R_{\rm vir}$) for \HI~$\lambda 1215$ (dark purple), \SiII~$\lambda 1260$ (light purple), \CII~$\lambda 1334$ (blue), \SiIII~$\lambda 1206$ (green), \CIII~$\lambda 977$ (yellow), \OVI~$\lambda 1031$ (pink), which is sorted by ionization potential. The shaded regions around each fit represent the $1~\sigma$ bootstrapped uncertainties.  Since 25 out of the 27 metal-line systems presented here reside inside the virial radius (see Table \ref{galaxy-abs}), we focus on the variation of {\fewcorot} within $1~R_{\rm vir}$. The dark purple dotted line presents the fit to the 39 {\HI} co-rotation fractions found within $1~R_{\rm vir}$ from \citetalias{GF1} for the purpose of direct comparison. On average, we observe a higher co-rotation in {\CII}, {\SiII}, {\SiIII} and {\CIII} lines closer to the galaxy, which tends to decrease with increasing $D/R_{\rm vir}$. It appears that the {\OVI} line is more constant across the full range of $D/R_{\rm vir}$ and the {\HI} trend is in between that of the low and high ions. To examine this further we explore the slopes of the fits to the $f_{\rm EWcorot}$ for each ion. 

Figure \ref{fig:DRvircorot}~(bottom) shows the slope of the corresponding fit to {\fewcorot} versus $D/R_{\rm vir}$ and the vertical error bars represent the $68\%$ confidence intervals obtained through bootstrapping. We find that all slopes are negative, but most are consistent with zero within uncertainties. We see hints that the slopes for the ions with lower ionization potential tend to be decreasing, while the ions with higher ionization potential (like {\OVI}) have a flatter slope, however we need more data to robustly determine this.

\subsection{Co-rotation of CGM phases vs azimuthal angle}
\label{3.3}

\begin{figure}
    \centering
    \includegraphics[width=\columnwidth]{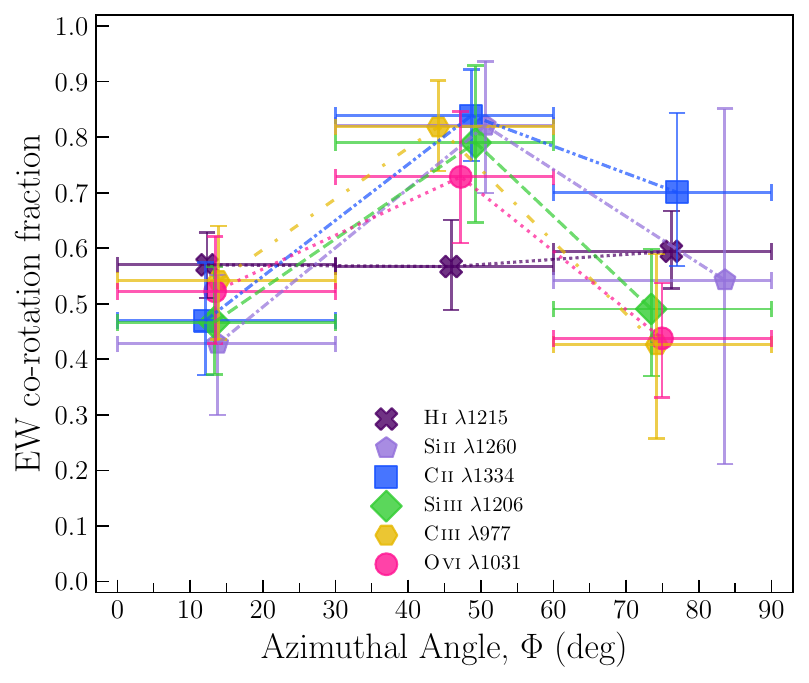}\hfill 
    \caption{Average equivalent width co-rotation fraction ($f_{\rm EWcorot}$) in 3 bins of azimuthal angles ($\Phi$). The distinct shapes corresponding to different ions are the averaged $f_{\rm EWcorot}$. The horizontal bars represent the azimuthal angles bins and the vertical error bars represent the $1\sigma$ bootstrapped errors on $f_{\rm EWcorot}$. The co-rotation fractions of all ions are nearly consistent with {\HI} within $1 R_{\rm vir}$, showing minimal variation across azimuthal angles with the errors, possibly peaking at the intermediate bin.}
    \label{fig:corot-AA}
\end{figure}
Previous studies have shown that  CGM gas exhibits a bimodality where the majority of the {\MgII} and {\OVI} were detected along the galaxies' major and minor axes \citep[e.g.,][]{Bordoloi2011, Bouche2012, Glenn2012, Glenn2015decmorpho, Lan2014, Dutta2017, Lundgren2021}. This is consistent with simulations where they depict the accretion of metal-poor gas along galaxies' major axis and enriched outflows perpendicular to the disk along the minor axis of the galaxies \citep{weiner2009,Nestor11,Bouche2012, Glenn14,Lan2018, Schroetter19_megaflow3}. In \citetalias{GF1}, the {\HI} co-rotation fraction has a constant average value across all azimuthal angles within $R_{\rm vir}$. However, beyond this region, it decreases by a factor of two for gas along the minor axis and for edge-on galaxies. This decrease along the minor axis was inferred to be due to outflows being bound to galaxy halos and not escaping. They further report that systems with significant co-rotation ($f_{\rm EWcorot} > 0.5 $) primarily exist along the major and minor axes of galaxies, where both inflows and outflows could have signatures of co-rotation.

\begin{figure}
    \centering
    \includegraphics[width=\columnwidth]{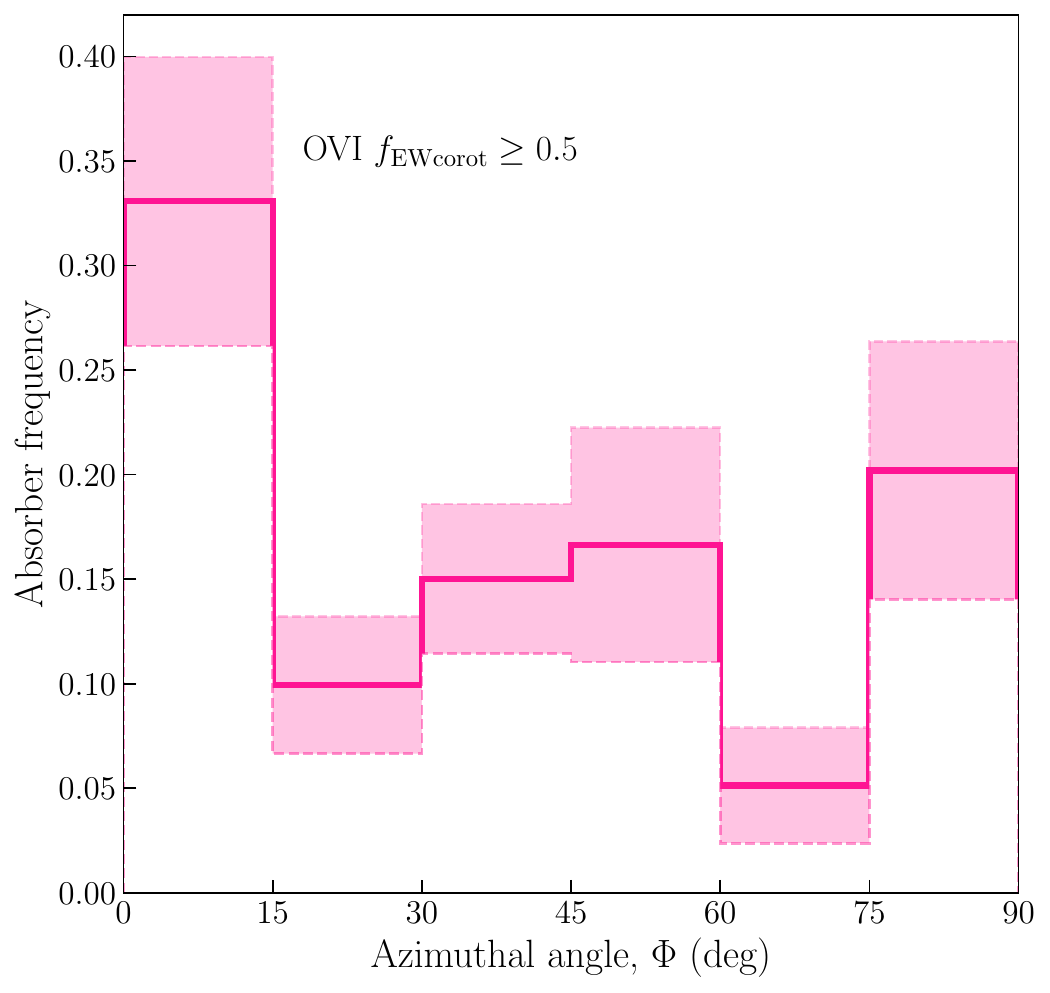}\hfill 
    \caption{Distribution of {\OVI} absorption systems that only have high co-rotation fractions (${\fewcorot \geq 0.5 }$) with respect to azimuthal angle. The observed frequency in each 15-degree bin of $\Phi$ is shown with the solid pink line, with the $1 \sigma$ error, measured by bootstrapping the sample, presented as the shaded region. Although we see a bimodality for the {\HI} {\fewcorot} (\citetalias{GF1}), here we only find high co-rotation fractions along the major axis. We note that we do not have enough statistics to test this with other ions.}
    \label{fig:OVI}
\end{figure}

Here we examine the connection between CGM gas flows in various ionization phases and their host galaxy orientations.
Figure~\ref{fig:corot-AA} presents $f_{\rm EWcorot}$ as a function of azimuthal angle ($\Phi$). Different point shapes represent the average {\fewcorot} in bins of $\Phi$, with different ions indicated as follows: dark purple crosses for {\HI}, light purple pentagons for {\SiII},  blue squares for {\CII}, green diamonds for {\SiIII}, yellow hexagons for {\CIII}, and pink circles for {\OVI}. The horizontal bars present the azimuthal angles bins while the vertical bars present the $1 \sigma$ bootstrap errors. For a fair comparison, we only plot 39 {\HI} systems from \citetalias{GF1} (purple distribution) detected within the virial radius of galaxies ($D/R_{\rm vir}\leq1$).

Similar to \citetalias{GF1}, we find that the co-rotation fractions of all ions are roughly consistent with the {\HI} as being nearly flat across all azimuthal angles with a possible peak at intermediate azimuthal angles. There is no variation between ions in each bin, within the errors shown here. Again, a larger sample may help to identify any possible differences if they exist. While this provides general statistics of the co-rotating gas as a function of azimuthal angle, what is more important is addressing where the co-rotating gas is primarily found. 

To address where the majority of co-rotating gas is found, we computed the frequency of absorption systems as a function of the azimuthal angle. However, given the sample sizes, we are only able to perform this analysis using {\OVI}, which has 15 galaxies hosting absorption with $f_{\rm EWcorot}\geq0.5$. We attempted this on other ions with no significant success as they have between 9--11 galaxies. Fig.~\ref{fig:OVI} shows only {\OVI} absorption systems that are dominated by co-rotating gas (e.g., the "bulk" of the gas where $f_{\rm EWcorot}\geq0.5$, which would be similar to other works). We adopt the methods of \citet{Glenn2012} and \citet{Glenn2015decmorpho} to create asymmetric univariate Gaussian PDFs \citep[see][]{kato2002} using the measured galaxy azimuthal angles and their asymmetric uncertainties. This results in the mean PDF of all galaxies as a function of $\Phi$, which represents the absorption frequency of co-rotating gas at a given $\Phi$. 

\begin{figure*}
    \centering
    \includegraphics[width=0.82\linewidth, height=22cm]{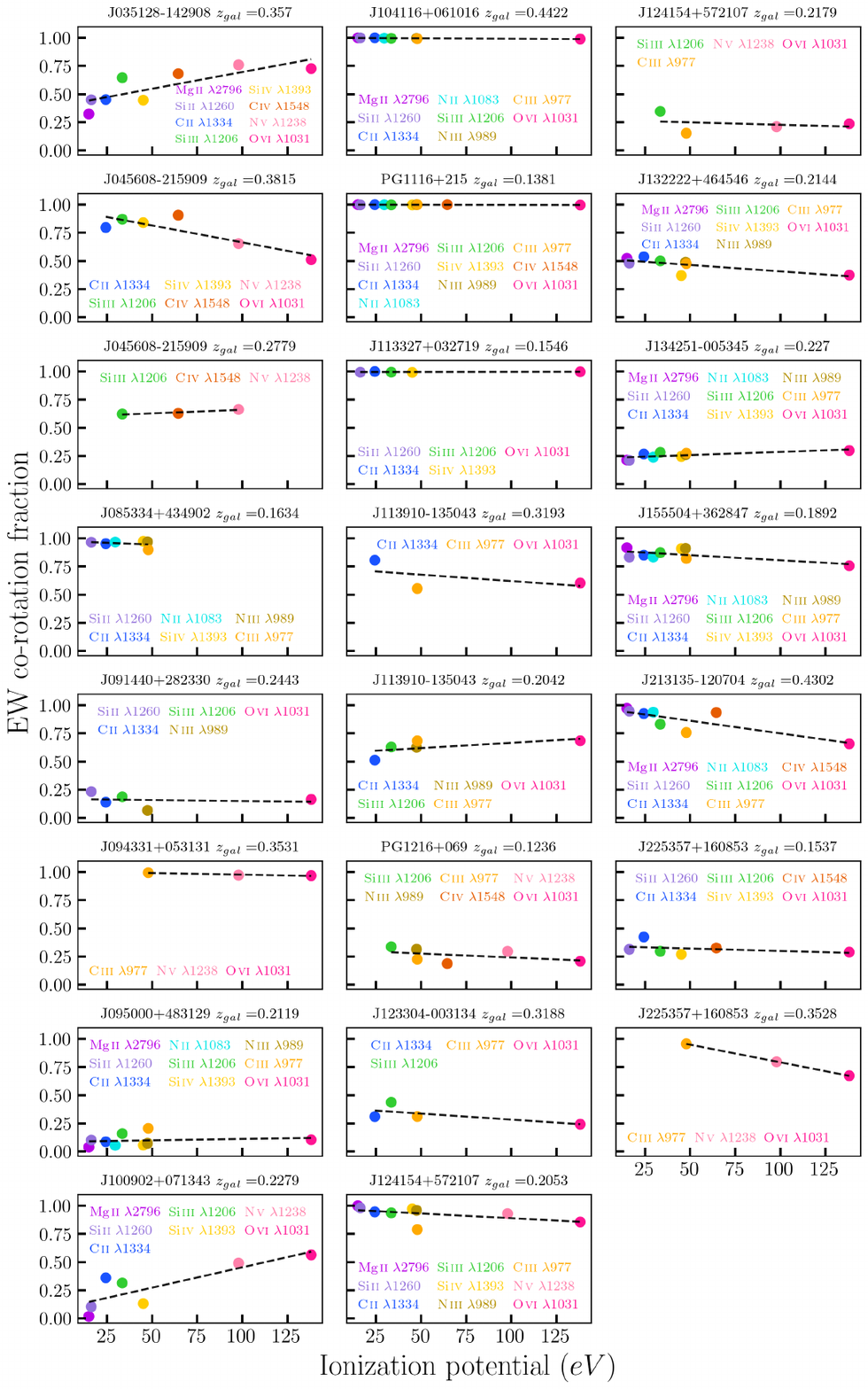}\hfill 
    \caption{Metal absorption lines detected in each system with at least 3 ions and their EW co-rotation fractions (\fewcorot) are displayed in each panel. Errors on {\fewcorot} are smaller than the points. The ions are colour-coded according to their ionization potential, with cooler colours representing the lower ionized species and warmer colours indicating the higher ionization states. The black dotted line presents the simple polynomial fit to the data points. The slope of the fit represents {\slope}, which is the change in the EW co-rotation fraction as a function of ionization potentials. The quasar field and galaxy redshift are labelled on top of each panel. The systems are arranged following the same order as Table~\ref{galaxy-abs} from the top left to the bottom right.  }
    \label{fig:slope_23sys}
\end{figure*}

Fig.~\ref{fig:OVI} shows the PDF, in 15~degree bins, for {\OVI} where the bulk of the absorption is consistent with a co-rotation model. We find that there is a significant peak within 15 degrees along the major axis of galaxies, where accretion is expected to occur. With our current dataset, we do not find a peak along the minor axis. As a verification, we ensured that our sample of {\OVI} absorbers has an azimuthal angle distribution consistent with being flat as expected for a random sample of galaxies. A Kolmogorov-Smirnov test shows that our sample is not discernible from a random population of galaxies within $1.3 \sigma$.

\subsection{Co-rotation fraction as a function of ionization}
\label{3.4}

Kinematic studies of the low-ionization CGM have found that {\MgII} absorption appears to be mostly consistent with the rotation of the host galaxies, while {\OVI} shows that only about half of the higher ionization-phase CGM exhibits signatures of co-rotation. These results could imply a difference between the fraction of gas that is co-rotating with galaxies in different ionization states. In Sections~\ref{3.1} and \ref{3.2}, we found that the CGM co-rotation fractions as a function of {\HI} column density and $D/R_{\rm vir}$ might be dependent on ionization. 

One of the primary objectives of this study is to investigate how the CGM absorption co-rotation fraction varies as a function of ionization. To address this we examine the {\fewcorot} of all the detected metals in each system, which is shown Figure~\ref{fig:slope_23sys}. Each panel of Figure~\ref{fig:slope_23sys} is for a different galaxy with the name and redshift labelled on the top. Each panel shows the co-rotation fraction for each ion with measurable absorption, and for galaxies that have at least three detected metal lines (23/27 galaxy--CGM pairs), with some galaxies having up to nine detected metal-lines. Their co-rotation fractions are shown as a function of ionization potential and the ions are colour-coded according to their ionization potential, with cooler colours representing the lower ionization species and warmer colours indicating the higher ionization species. The legend in each panel presents the colours and names corresponding to detected ions. The dashed line is a first order polynomial fit to the data, where we have measured the change in the EW co-rotation fraction as a function of ionization ({\slope}) and $1\sigma$ errors. 

From Figure~\ref{fig:slope_23sys}, we identify that while each CGM ion has a range of co-rotation fractions, the co-rotation in individual systems also varies with ionization. In some cases, the range in {\fewcorot} for an individual system can vary by as much as a factor of five (e.g., J100902$+$071343, $z_{\rm gal}=0.2279$) or show little variation at all (e.g., J104116$+$061016, $z_{\rm gal}=0.4422$), which indicates a wide distribution of the changes in the co-rotation fraction as a function of ionization (\slope). The {\slope} in this distribution can also be positive or negative.  

\begin{figure}
    \centering
    \includegraphics[width=\columnwidth]{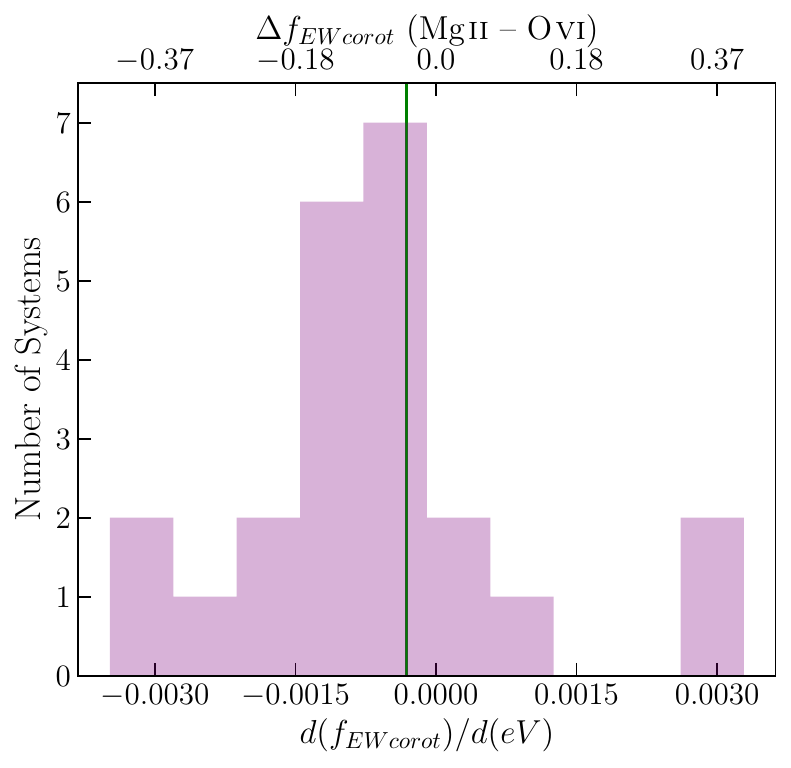}\hfill 
    \caption{The distribution of the change in the co-rotation fraction as a function of ionization potential ({\slope}) for the fits to the EW co-rotation fractions as a function of ionization potential shown in Figure~\ref{fig:slope_23sys}. The green line represents the average of the bootstrapped {\slope}=$-0.00032\pm0.00001$. This indicates that the majority of the systems exhibit negative {\slope}, implying that, on average, low-ionization gas has a higher co-rotation fraction compared to the higher-ionization phase. This is reflected in the top axis, which shows the computed difference between co-rotation fractions {\MgII} of and {\OVI} (\dfcorot), which covers our full range of ionization potentials.}
    \label{fig:slope_hist}
\end{figure}

Figure~\ref{fig:slope_hist} presents a histogram of {\slope} from Figure~\ref{fig:slope_23sys}. The vertical green line shows the variance-weighted mean of the bootstrapped distribution while including their $1~\sigma$ standard deviations in the re-sampling. The variance-weighted mean  of ${\slope}=-0.00032\pm0.00001$, which indicates that we primarily find systems with negative ${\slope}$. This histogram indicates that the majority of the systems ($16/23=70\%$), have higher co-rotation fractions in the low-ionization CGM compared to the high-ionization CGM. 

To better quantify the co-rotation fraction of changes between different phases, Figure~\ref{fig:slope_hist} also shows the difference between co-rotation fractions of the ion with the lowest ionization energy ({\MgII}) and the ion with the highest ionization energy ({\OVI}), which is defined as {\dfcorot}. We do note that the {\MgII} and/or {\OVI} were not detected in all of the systems as shown in Table \ref{EW-detections}. Thus we used {\slope} to compute the {\dfcorot}, which represents the maximum change in the co-rotation fraction for low to high ionization species.
The {\dfcorot} ranges between $-0.4$ and $0.4$, with most of the systems residing between $0$ and $-0.2$ with a median value of $-0.05$. This implies that while the majority of systems exhibit a higher co-rotation fraction for lower-ionization species compared to higher-ionization species, the difference can be small and suggests that there is significant kinematic consistency between ions. 

\begin{figure}
    \centering
    \includegraphics[width=1\columnwidth]{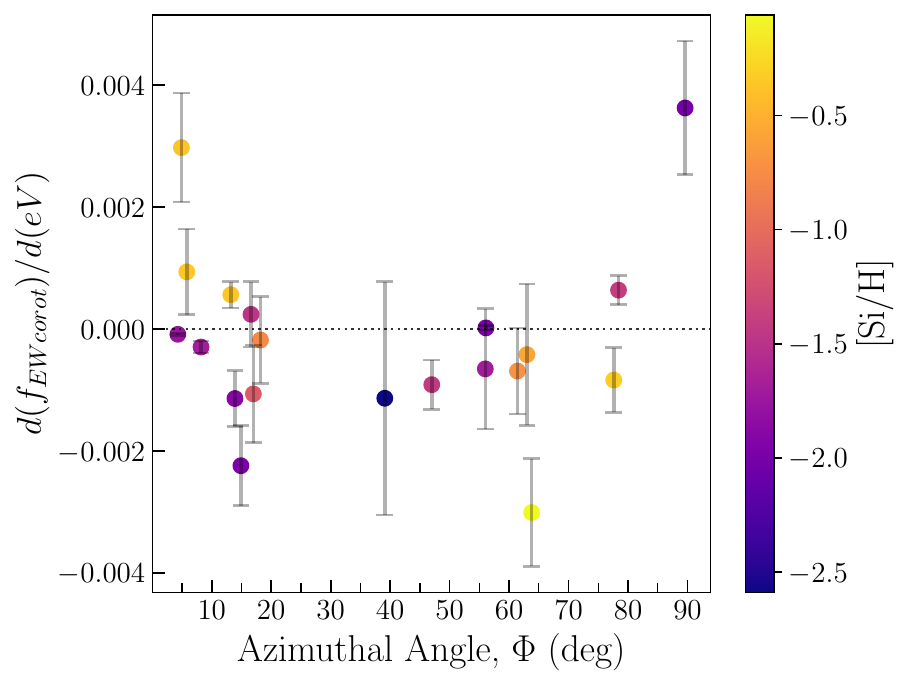}\hfill
    \caption{{\slope} as a function of azimuthal angle. The vertical bars represent $1\sigma$ errors, and the data points are colour-coded based on the CGM metallicities adopted from \citep{pointon19}. We find that the majority of the systems have negative {\slope}, which implies that, on average, ions with lower ionization potential have higher co-rotation fractions compared to ions in higher CGM phases. Also, there are more systems with positive {\slope} along the major and minor axes. This could imply that on average, higher ionization species have higher co-rotation fractions in these two preferential spatial projections.  We further note that metallicity increases with increasing {\slope} along the major axis, while the metallicity decreases with increasing {\slope} along the minor axis. }
    \label{fig:slope-AA}
\end{figure}

In order to understand the origins of the different values of {\slope}, we explored how they vary with galaxy properties such as halo mass, inclination and azimuthal angles, ISM and CGM metallicity, {\HI} column density, and the impact parameter. Some trends might be expected given that ionization may change as a function of distance from the host galaxy, within outflows, or for different {\HI} column densities. However, we did not find any significant connection between {\slope} and the aforementioned galaxy and/or gas properties with our current sample size. We do find a hint of a trend, which we describe in Figure~\ref{fig:slope-AA}. Figure~\ref{fig:slope-AA} shows {\slope} as a function of azimuthal angle. Here we find that positive {\slope} values seem to reside around the major and minor axes of galaxies. Although this is only for a few systems, it is interesting that they do exist in those preferential locations.  

Figure~\ref{fig:slope-AA} also shows {\slope} as a function of azimuthal angles that are colour-coded with CGM metallicities ([Si/H] ratio). We adopted the CGM metallicities that were available for 20 systems in our sample from \citet{pointon19}. To derive the CGM metallicites, \citet{pointon19} implemented the Cloudy ionization modelling software \citep{ferland13_cloudy} and the HM05 UV background \citep{HaardtMadau1996} to generate grids of ionization properties and fit the grids. They employed the Markov Chain Monte Carlo (MCMC) technique, following the procedure described by \citet{crighton2013b}, to ascertain the most probable metallicity, represented as the [Si/H] ratio.

We find that within 20 degrees of the galaxies' major axes, systems with smaller {\slope} exhibit lower metallicity (darker purple), whereas systems with larger {\slope} exhibit higher metallicity. Having fewer data points, we do see that for absorption along the minor axis ($\Phi \geq 60$ degrees), the opposite is true. In fact, systems with smaller {\slope} exhibit higher metallicity (yellow), whereas systems with larger {\slope} exhibit lower metallicity (darker purple). Overall,this may suggest a potential connection between gas accretion and CGM metallicity, particularly along the major and minor axes, where we expect gas accretion and outflows, respectively, to occur. We discuss this further in Section~\ref{sec:discussion}.


\begin{figure*}
    \centering
    \includegraphics[width=0.9\linewidth]{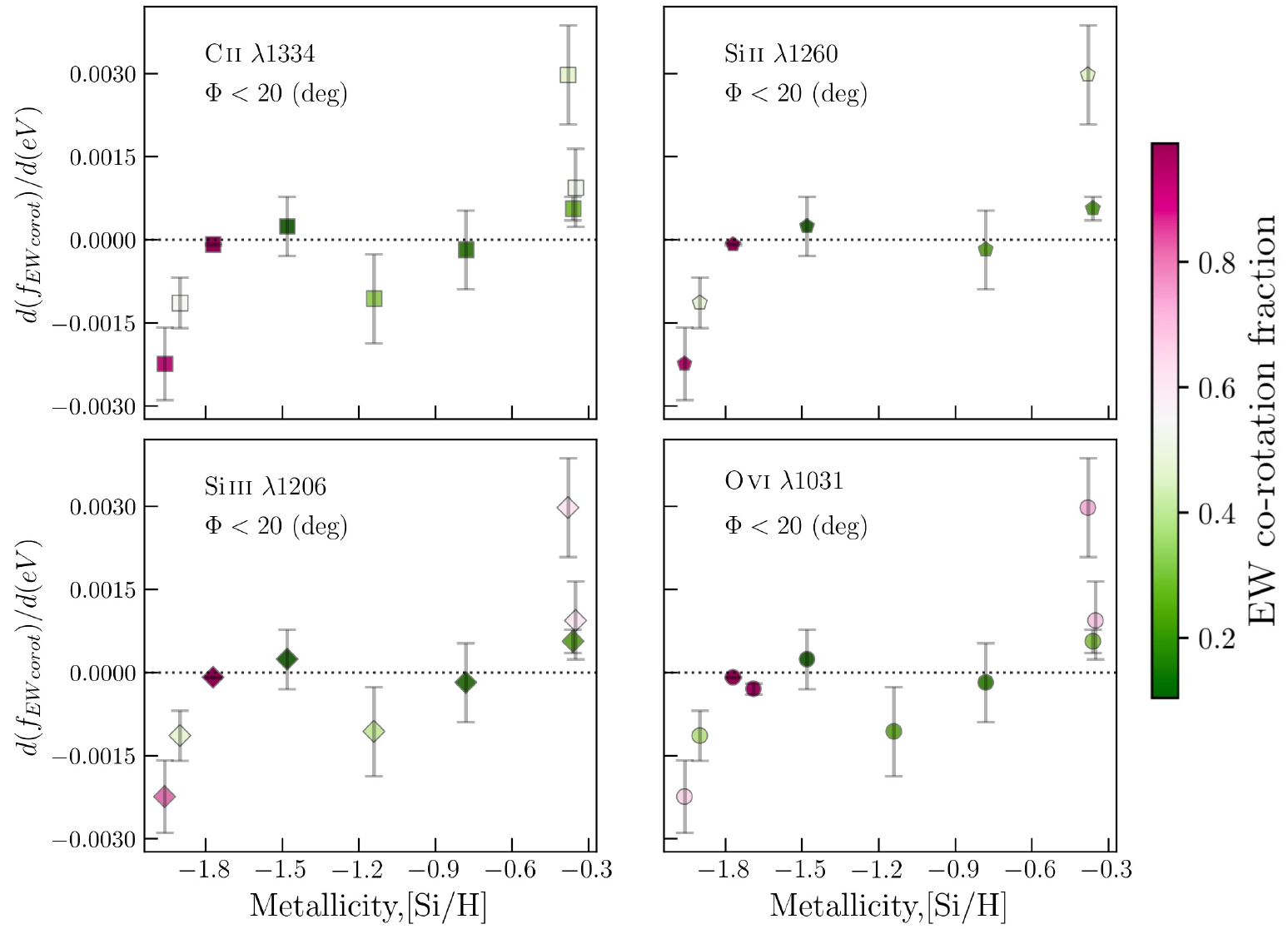}\hfill 
    \caption{The change in the EW co-rotation fraction as a function of ionization ({\slope}) versus metallicity [Si/H] for systems within $20$~degree of the major axis.
    Points are colour-coded by EW co-rotation fraction of each ion. Comparing the four panels, we find that the major axis absorption with lower ionization energy and higher co-rotation fraction is associated with lower metallicities. While absorption with higher ionization energy shows high co-rotation fractions at the highest metallicities. This could be suggestive of two forms of accretion: 1) low ionization gas that is co-rotating with the galaxy having the lowest metallicity, which is accreting along cosmic filaments 2) high ionization and higher metallicity gas being recycled material that is cooling and joining with the accretion along the major axis. See Section~\ref{4.2} for further discussion. } 
    \label{fig:slope-metal}
\end{figure*}

\section{Discussion}
\label{sec:discussion}


Studying how the CGM is kinematically coupled to its galaxy is critical in addressing how baryons are being processed. As the CGM is multiphase, it is unclear whether the different gas phases probe different baryon cycle processes.  Here we investigate the galaxy--multiphase CGM kinematic connection primarily via {\SiII}, {\CII}, {\SiIII}, {\CIII} and {\OVI} to determine the statistical nature of how different phases are consistent with a co-rotation/accretion scenario.  


\subsection{\boldmath{\fewcorot} dependence on N({\HI}) and \boldmath$D/R_{\rm vir}$}

Simulations show that the CGM gas with higher {\HI} column density is more likely associated with accretion and outflows \citep{Fumagalli2011, suresh2019}.
From observations, \cite{Cote2005} found that the lower column density {\HI} gas is likely connected to IGM filaments and with little-to-no co-rotation signatures. \citetalias{GF1} also investigated the {\HI} CGM kinematics with respect to the galaxy rotation and found a correlation between the {\HI} column density and its co-rotation with the host galaxy. 
Our findings are consistent with previous studies and show  that for systems associated with higher column density {\HI}, low-ionization absorbers are more kinematically consistent with their host galaxies compared to high-ionization absorption. This could imply that low-ionization species are likely a better tracer of gas flows. However, given that N$({\HI})$ is anti-correlated with galaxy--quasar projected separation\citep[and references therein][]{Glenn2021}, it is important to address how the co-rotation fraction varies with the impact parameter.


\cite{French2020} reported a decrease in {\Lya} co-rotation fraction with increasing impact parameter. However, their sample mostly included absorbers with low column densities. \citetalias{GF1} used a larger sample of {\HI} absorption, with a wider range in column density, and normalised the impact parameters to the virial radius to account for the different halo masses and the self-similarity of the CGM \citep{Churchil13_selfsimilar}.
Their results demonstrated a slight increase in the distribution of co-rotation fractions within the virial radius, particularly for high column density systems. However, beyond the virial radius, where lower column density systems dominate, co-rotation fractions decreased with increasing $D/R_{\rm vir}$.

\citet{Klimenko23} presented a kinematics comparison between six galaxies and their {\HI}$+$metal CGM absorption.  They found agreement between the velocities of the strongest {\HI} absorption components and the predicted radial velocities of the galactic disks. This occurs with 10 galaxy effective radii ($\sim 60$~kpc), where absorption beyond this distance is mostly in the opposite direction to that expected from the galactic disk rotation. These results are consistent with a decreasing co-rotation fraction with increasing distance.

As shown in Section~\ref{3.1}, our results suggest an anti-correlation between the EW co-rotation fraction of different ions and the $D/R_{\rm vir}$. While the {\fewcorot} of species with lower ionization energy may gradually decrease with increasing $D/R_{\rm vir}$, the {\fewcorot} of species with higher ionization energy remains almost constant across $D/R_{\rm vir}$. Our findings indicated that closer to the galaxies, the low-ionization absorbers tend to have a higher EW co-rotation fraction compared to high-ionization absorption. This could provide further evidence for our expectation of low-ionization CGM being more associated with accretion scenario \citep[e.g.,][]{Glenn2010a, Ho17, MartinChris2019} compared to high-ionization CGM \citep{Glenn2019}.
This could also be consistent with the simulations of \citet{oppenheimer2018} who showed that for the hot gas, the tangential (line of sight) velocity versus $D/R_{\rm vir}$ trend is less steep than that of the cool gas. This could imply that the kinematics of the cool gas in CGM varies less significantly with the impact parameter normalized by virial radius and the cool gas can likely be more coupled to the host galaxy.

It might be interesting to note that on average, the dependence of {\fewcorot} on $D/R_{\rm vir}$ is quite similar for all ions. This could imply that {\OVI} is photoionized and remains so at all distances. This may differ from simulations that suggest we may find more photoionized {\OVI} at only larger $D/R_{\rm vir}$, especially at higher redshifts \citep{strawn2021}.

Figure~\ref{fig:DRvircorot} and Figure~\ref{fig:NHIcorot} shows that for systems closer to galaxies (lower $D/R_{\rm vir}$), or associated with higher {\HI} column density, the average {\fewcorot} of {\HI} appears to be in between the low and high ionization absorbers. This could be due to the {\HI} contributing to both gas phases.
Although, the slope of the EW co-rotation fraction versus N$({\HI})$ is positive for all the ions, the trend for high-ionization absorption seems to be flat within the error. Thus for a further understanding of the change in the trends as a function of ionization, one would also have to consider how much {\HI} is associated with each phase, i.e., {\OVI} may only be associated with some fraction of the total {\HI} \citep[e.g.,][]{muzahid15,sameer2021, Hasti2021}. 






\begin{figure}
    \centering
    \includegraphics[width=1\columnwidth]{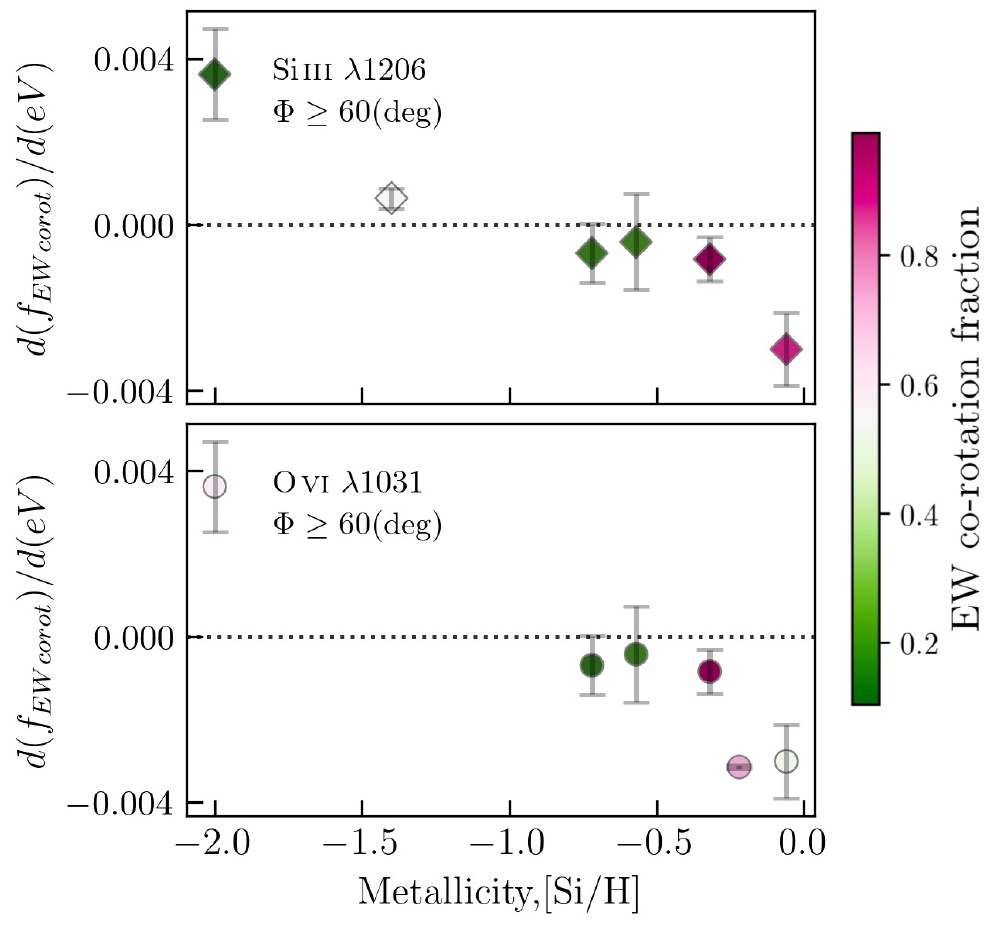}\hfill 
    \caption{The change in the EW co-rotation fraction as a function of ionization potential ({\slope}) versus metallicity [Si/H] for systems with $\Phi \geq 60$~degree (along the minor axis). Points are colour coded by EW co-rotation fraction for {\SiIII} (top) and {\OVI} (bottom). Minor axis absorbers, either with low or high ionization energy mostly exhibit negative {\slope} values. 
    Opposite to major axis absorption, systems with negative {\slope} have higher metallicities. Systems with higher metallicity and also higher co-rotation fractions (white to pink shades) could potentially indicate metal-rich outflows that are still kinematically connected to the disk where they are ejected from. The minor axis systems, with low co-rotation fractions could be interpreted as the recycled gas that may become kinematically coupled and eventually accrete onto the galaxy.
 } 
    \label{fig:slope-metal_minorax}
\end{figure}


\subsection{\boldmath{\fewcorot} dependence on \boldmath$\Phi$ and metallicity}

\label{4.2}

Previous observations have shown that the distribution of the CGM absorption is not uniform; instead, it exhibits a bimodality where the majority of the {\MgII} and {\OVI} were detected along the galaxies' major and minor axes \citep[e.g.,][]{Bordoloi2011, Bouche2012, Glenn2012, Glenn2015decmorpho, Lan2014, Dutta2017, Lundgren2021}. This is consistent with simulations where they depict the accretion of metal-poor gas along galaxies' disks and enriched outflows perpendicular to the disk and along the galaxy minor axis \citep{weiner2009,Nestor11,Bouche2012, Glenn14,Lan2018, Schroetter19_megaflow3}. 

We find that the co-rotation fractions of all ions are roughly consistent with the {\HI}, remaining nearly flat across all azimuthal angles, with a potential peak at intermediate azimuthal angles. 
However, we next ask where gas resides that has the bulk (>50\%) of the absorption consistent with co-rotation. while this could not be done with all ions, we find that the {\OVI} residing along the major axis of galaxies tends to be co-rotating (Figure~\ref{fig:OVI}).  This differs from \citetalias{GF1}  where they found that the {\HI} has most of the co-rotating gas along the major and minor axes. It is unclear why this occurs as one would expect {\OVI} to also occur in outflows, but maybe has little co-rotation signatures. It is possible that the low-ionization gas mimics the {\HI} bimodality in the co-rotation fraction with azimuthal angle, however we are unable to test this with our current sample size. Although given that all metal ions seem to follow the same trend, it is not implausible that they exhibit a similar distribution to {\OVI}; more studies are required here.

Despite this bimodal spatial distribution of the CGM, it is still debated how CGM metallicity varies with azimuthal angle. We might anticipate that gas with lower metal content would tend to accrete along the galaxy's disk. Conversely, one would expect metal-enriched outflows to be more prevalent along the minor axis. This simple picture of the CGM is also consistent with predictions from simulations \citep[e.g.,][]{nelson2019, peroux2020b,Weng2023}. Yet observations of CGM metallicity \citep[][]{peroux2016, pointon19, Glenn2019b_metal, sameer2024} do not reveal a dependence on geometry. However, when dust is accounted for in the metallicity calculation, it was determined that lower metallicity can be found along the major axis and a higher metallicity along the minor axis of galaxies \citep{Wendt2021}.

Although our results do not imply a significant relation between the {\fewcorot} of different ions and their azimuthal distributions, there are some hints of a connection between the {\fewcorot}--ionization slope ({\slope}) and the major and minor axes systems' metallicity (Figure~\ref{fig:slope-AA}). Figure~\ref{fig:slope-metal} shows only systems residing along the galaxies' major axis (within 20 degrees) and their {\slope} as a function of CGM metallicity. The upper left panel shows the systems with detected {\CII} absorption and the data points are colour-coded based on its {\fewcorot}. The upper right panel shows the same but for {\SiII}. Although the sample is small, we find that for low ionization gas (both {\CII} and 
{\SiII}), the lowest metallicity gas has the highest co-rotation fraction. The absorbers with the lowest metallicities also have negative {\slope}, which indicates cooler, less ionized gas has a higher consistency with a co-rotation model than the higher ionization gas. On the contrary, absorption systems with only a small fraction of absorption consistent with co-rotation have higher metallicities and {\slope} that are consistent with being positive. It would seem like both metallicity and co-rotation fraction could be an effective way of identifying cosmological accretion onto galaxies.  The bottom panels of Figure~\ref{fig:slope-metal} are the same as the top but for {\SiIII} and {\OVI} absorption.  While we find that both {\SiIII} and {\OVI} have high EW co-rotation fractions at low metallicities, they also have high EW co-rotation fractions for the highest metallicity systems, which is not seen in the lower phase. 

Overall these results are intriguing and could be suggestive of two forms of accretion. The low ionization gas that is co-rotating strongly with the galaxy has the lowest metallicity, which could arise from accretion along cosmological filaments containing metal-poor gas that is driving the angular momentum of their galaxies. The signature of co-rotation is also seen in the higher ionization phases also predicted by simulations \citep{oppenheimer2018,stern2023,Hafen2022}, which indicates that it is a multiphase accretion. On the other hand, the higher ionization lines also show high co-rotation fractions (\fewcorot$>0.5$) for the highest metallicity systems in our sample. It is unlikely that this is the edge of the filaments or a higher ionization sheath that surrounds the filaments given the higher metallicity. One possibility is that this gas is recycled material that is cooling and joining with the accretion along the major axis. This is consistent with cosmological simulations predictions of hot gas ($T\sim10^{6}$~K) that accretes through the CGM and cools as it joins the extended plane of the disk \citep{stern2023}. Using FIRE simulations, \citet{Hafen2022} predicted that the rotating cooling gas is the dominant source of accretion onto Milky-Way mass galaxies. However, the hot gas at virial temperature decelerates with a rotation pattern and then starts to cool down and accretes onto the disk. Thus, we could be observing two different phases of gas accretion. This may also be indicative of fountain accretion \citep{Fraternali2017} at larger scales and distances from the disk. As the feedback materials with higher metallicity travel through the halo and mix with the hot CGM, it triggers and accelerates the cooling of hot gas which leads to the accretion of gas from the hot phase of CGM. 

Figure~\ref{fig:slope-metal_minorax} presents {\slope} as a function of CGM metallicity for systems observed along the minor axis of galaxies ($\Phi \geq 60$~degree). While we have even fewer data, and fewer ions, along the galaxy minor axis, we still find it interesting to discuss and hope to expand on the sample in future studies. The top panel shows the systems with detected {\SiIII} absorption and the data points are colour-coded based on its {\fewcorot}. The bottom panel show the same but for {\OVI} absorption. We find that minor axis absorbers, mostly have negative {\slope} values and exhibit higher metallicities, which is true for both low-ionization {\SiIII} absorption and high-ionization {\OVI}. This is the opposite trend compare to major axis systems presented in Figure~\ref{fig:slope-metal}, where systems with smaller {\slope} and higher EW co-rotation fraction are associated with the highest metallicities. This could be explained as outflowing material that is still kinematically coupled to the host galaxy. This is consistent with \citetalias{GF1}, where we surprisingly found the signatures of co-rotation in {\HI} absorption along the minor axis within the virial radius of galaxies. Whereas, systems with smaller {\slope} and low EW co-rotation fraction could be explained as recycled gas that is flowing back onto the galaxy.

Regardless of the underlying spatial, kinematic, and ionization structure of the CGM inferred in this discussion, the co-rotation fractions provide stringent constraints on simulations. They have proven to be a valuable quantifier for characterizing the azimuthal, metallicity, and kinematic relationships of the CGM, challenging simulators and inviting them to adopt these quantifiers.

\section{Conclusion}
\label{sec:conslusion}

We used the EW co-rotation fraction (\fewcorot) method that we developed in \citetalias{GF1} to study the kinematic connection between the multiphase CGM absorption and host galaxy rotation. Here we examined 27 galaxy--CGM pairs with galaxies span over a redshift range of 0.09--0.5.  When possible, we use the CGM metal absorption lines {\CII~$\lambda 1334$, \CIII~$\lambda 977$, \CIV~$\lambda 1548$, \SiII~$\lambda 1260$, \SiIII~$\lambda 1206$, \SiIV~$\lambda 1393$, \NII~$\lambda 1083$, \NIII~$\lambda 989$, \NV~$\lambda 1238$, \MgII~$\lambda 2796$, and \OVI~$\lambda 1031$}, to study the {\fewcorot} for different ions.
The absorption lines are detected in the spectra of background quasars ($D=21-276$~kpc) observed with {\it HST}/COS and Keck/HIRES or VLT/UVES. 
The galaxies have a halo mass range of $\log (M_{\rm h}/M_{\odot})=10.5-12.3$ and their rotation curves were obtained with Keck/ESI as part of the analysis in \citetalias{GF1}. 
We investigate the dependence of {\fewcorot} on {\HI} column density (\colden), impact parameter normalised to the virial radius, and azimuthal angle. We examine the change in the EW co-rotation fraction as a function of ionization potential ({\slope}) and also explore any connection between the distribution of {\slope} as a function of galaxy and gas properties. We summarise our results as follows:

\begin{enumerate}
    \item The median EW co-rotation fractions for all the ions are consistent within the errors with {\SiII}: $0.5\pm0.4$, {\CII}: $0.8\pm0.3$, {\SiIII}: $0.5\pm0.3$, {\CIII}: $0.7\pm0.3$ and {\OVI}: $0.6\pm 0.3$. Thus, for a general population of absorbers, our results suggest that all ions appear to have similar amounts of co-rotation.

    \item We find that the EW co-rotation fraction of lower- and higher ionization potential species are likely increasing with {\colden}. This correlation is likely shallower for higher ionization gas.

    \item The {\fewcorot} of {\CII}, {\SiII}, {\SiIII} and {\CIII} decreases with increasing $D/R_{\rm vir}$ while the {\OVI} co-rotation is more constant across the full range of $D/R_{\rm vir}$, which is similar to {\HI}, and {\HI} could be flatter since it traces both low and high ions. 

    \item The average {\fewcorot} of all the ions is consistent with almost being flat across all $\Phi$ bins within the errors, however there is a possible peak at intermediate azimuthal angles. We do not find a significant variation between different ions in each $\Phi$ bin.

    \item The larger number of {\OVI} detections enabled us to investigate where the majority of co-rotating gas is found. Highly co-rotating {\OVI} primarily resides along the galaxies' major axis. This differs from the results in \citetalias{GF1}, where they showed a bimodal distribution of co-rotating {\HI} systems within the CGM with respect to azimuthal angles. 

    \item We examine the change in the EW co-rotation fraction as a function of ionization potential ({\slope}) for individual systems. We find that on average, the co-rotation fraction decreases with increasing ionization potential. Our results demonstrate that the low-ionization CGM has a higher co-rotation fraction compared to the high-ionization CGM. 

    \item Our results suggest a dependence of {\slope} on CGM metallicity. It appears that within 20 degrees of the galaxies’ major axes ($\Phi<20$), systems with smaller {\slope} have lower metallicities and high co-rotation fractions (consistent with cosmic filament accretion), whereas systems with larger slopes exhibit higher metallicities and high co-rotation fractions (which could be recycled gas being accreted). On the other hand, along the minor axis ($\Phi \geq 60$),  absorbers have negative {\slope} with higher metallicities consistent with outflows (for those having high co-rotation fraction) or recycled gas (for those having lower co-rotation fractions).

\end{enumerate}

In this work, for the first time, we investigate how the kinematics of the multiphase CGM absorption relates to galaxy rotation. Our results imply a significant amount of gas that is consistent with co-rotation in both low- and high- ionization CGM.  By examining the change in co-rotation as a function of ionization potential, we demonstrated that the low-ionization phase is more kinematically consistent with the host galaxy rotation and thus an accretion scenario. The evidence of co-rotation that we found in high-ionization gas could imply the hot CGM accretion. While there is a lot to be learnt from examining single ions and their connections to galaxies, as we have shown here, having a large range of ions allows us to reveal the underlying relationship between galaxies and their multiphase gas flows. While we find evidence for a relationship between the spatial distribution of the gas, its kinematics, and metallicities, we require additional data to fully quantify how gas accretion, both for low and high ionization metal lines, occurs onto galaxies.

\section*{Acknowledgements}
The authors thank the referee for insightful comments that have improved the manuscript. H.N, G.G.K, and N.M.N.\ acknowledge the support of the Australian Research Council Centre of Excellence for All Sky Astrophysics in 3 Dimensions (ASTRO 3D), through project number CE170100013. M.T.M acknowledges the support of the Australian Research Council through Future Fellowship grant FT180100194. Some of the data presented herein were obtained at the W.~M.~Keck Observatory, which is operated as a scientific partnership among the California Institute of Technology, the University of California and the National Aeronautics and Space Administration. The Observatory was made possible by the generous financial support of the W.~M.~Keck Foundation. Observations were supported by Swinburne Keck programs with 2010A\_W007E, 2010B\_W032E, 2014A\_W178E, 2014B\_W018E, 2015\_W187E, and 2016A\_W056E. The authors wish to recognise and acknowledge the very significant cultural role and reverence that the summit of Maunakea has always had within the indigenous Hawaiian community. We are most fortunate to have the opportunity to conduct observations from this mountain.

\section*{Data Availability}

The data underlying this paper will be shared following mutually agreeable arrangements with the corresponding authors.


\bibliographystyle{mnras}
\bibliography{paper3}



\appendix

\section{Galaxy--CGM absorption observation}

Details of each galaxy--absorption pairs are provided in Table \ref{galaxy-abs}. The table presents the background quasar field, galaxy ID, galaxy coordinates and redshift ($z_{\rm gal}$), impact parameter ($D$) and virial radius normalised impact parameter ($D/R_{\rm vir}$), galaxy inclination angle ($i$), the azimuthal angle ($\Phi$), neutral hydrogen column density (\colden) and halo mass ($\log (M_{\rm h}/M_{\odot})$).

\begin{landscape}
\begin{table}
\centering
\renewcommand{\arraystretch}{1.6}
\caption{Galaxy--absorption observations.}
\label{galaxy-abs}

\begin{threeparttable}
\begin{tabular}{l l c c c c c c c c c c}
\hline
\hline

  \multicolumn{1}{l}{Quasar} &
  \multicolumn{1}{l}{Galaxy} &
  \multicolumn{1}{c}{RA$_{\rm gal}$} &
  \multicolumn{1}{c}{DEC$_{\rm gal}$} &
  \multicolumn{1}{c}{$z_{\rm gal}$} &
  \multicolumn{1}{c}{$D$ (kpc)} &
  \multicolumn{1}{c}{$D/R_{\rm vir}$} &
  \multicolumn{1}{c}{$i$ (deg)} &
  \multicolumn{1}{c}{$\Phi$ (deg)} &
  \multicolumn{1}{c}{\colden} &
  \multicolumn{1}{c}{log($M_{\rm h}/M_{\odot})$} &
  \multicolumn{1}{c}{$g-r$}\\ 

\hline
  J035128$-$142908 & J0351G1 & 03:51:27.87 & $-$14:28:57.9 & 0.356992 & 72.3$\pm$0.4 & $0.53\substack{+0.07 \\ -0.07}$ & $28.5\substack{+19.8\\ -12.5}$ & $4.9\substack{+33 \\ -4.9}$& 16.86 $\pm$ 0.03 & $11.55\substack{+0.1 \\ -0.1}$ & 0.29\\ 
  J040748$-$121136 & J0407G1 & 04:07:49.67 & $-$12:11:05.5 & 0.495164 & 107.6$\pm$0.4  & $0.78\substack{+0.07 \\ -0.07}$ & $67.2\substack{+ 7.6\\ -7.5}$ & $21.0\substack{+5.3 \\ -3.7}$ & 14.34 $\pm$ 0.56 & $11.54\substack{+0.1 \\ -0.1}$ & 0.45\tnote{a}\\  
  J045608$-$215909 & J0456G1 & 04:56:08.93 & $-$21:59:29.2 & 0.381511 & 103.4$\pm$0.3 & $0.6\substack{+0.1 \\ -0.07}$ & $57.1\substack{+ 19.9\\ -2.4}$ & $63.8\substack{+4.3 \\ -2.7}$ & 15.1 $\pm$ 0.39 & $11.86\substack{+0.12 \\ -0.1}$ & 0.45\tnote{a}\\
  J045608$-$215909 & J0456G2 & 04:56:09.69 & $-$21:59:03.9 & 0.277938 & 50.7$\pm$0.4 & $0.4\substack{ +0.07 \\ -0.07 }$ & $71.2\substack{ +2.2 \\ -2.6 }$ & $78.4\substack{ +2.1 \\ -2.04 }$ & 14.78 $\pm$ 0.22 & $11.49\substack{ +0.1 \\ -0.1 }$ & 0.45\tnote{a}\\
  J085334$+$434902 & J0853G2 & 08:53:45.24 & $+$43:51:08.2 & 0.163403 & 26.2$\pm$0.1 & $0.18\substack{ +0.07 \\ -0.07 }$ & $70.1\substack{ +1.4 \\ -0.8 }$ & $56.0\substack{ +0.8 \\ -0.8 }$ & 19.93 $\pm$ 0.01 & $11.7\substack{ +0.1 \\ -0.09 }$ & 0.45\tnote{a}\\
  J091440$+$282330 & J0914G1 & 09:14:41.76 & $+$28:23:51.2 & 0.244312 & 105.9$\pm$0.1 & $0.81\substack{ +0.07 \\ -0.07 }$ & $39.0\substack{ +0.4 \\ -0.2 }$ & $18.2\substack{ +1.1 \\ -1.0 }$ & 15.55 $\pm$ 0.03 & $11.54\substack{ +0.1 \\ -0.1 }$ & 0.17\\
  J094331$+$053131 & J0943G1 & 09:43:30.72 & $+$05:31:17.5 & 0.353052 & 96.5$\pm$0.3 & $0.78\substack{ +0.07 \\ -0.07 }$ & $44.4\substack{+ 1.1 \\ -1.2 }$ & $8.2\substack{ +3.0 \\ -5.0 }$ & 16.46 $\pm$ 0.03 & $11.44\substack{ +0.1 \\ -0.1 }$ & 0.29\\
  J095000$+$483129 & J0950G1 & 09:50:01.01 & $+$48:31:02.3 & 0.211866 & 93.6$\pm$0.2 & $0.43\substack{ +0.2 \\ -0.12 }$ & $47.7\substack{ +0.1 \\ -0.1 }$ & $16.6\substack{ +0.1  \\ -0.1 }$ & 18.48 $\pm$ 0.19 & $12.22\substack{+ 0.24 \\ -0.17 }$ & 0.45\tnote{a}\\
  PG1001$+$291 & PG1001G1 & 10:04:02.37 & $+$28:55:12.3 & 0.137403 & 56.7$\pm$0.2 & $0.93\substack{+ 0.06 \\ -0.07 }$ & $79.1\substack{+2.2  \\ -2.1 }$ & $12.4\substack{ +2.4 \\ -2.9 }$ & 14.98 $\pm$ 0.03 & $10.59\substack{+ 0.08 \\ -0.1 }$ &  0.20\\
  J100902$+$071343 & J1009G1 & 10:09:02.74 & $+$07:13:37.7 & 0.227855 & 64.0$\pm$0.8 & $0.44\substack{+ 0.08 \\ -0.07 }$ & $66.3\substack{+ 0.6 \\ -0.9 }$ & $89.6\substack{+ 0.4 \\ -1.3 }$ & 17.23 $\pm$ 0.16 & $11.68\substack{+ 0.1 \\ -0.09 }$ & 0.45\tnote{a}\\
  J104116$+$061016 & J1041G1 & 10:41:06.32 & $+$06:09:13.5 & 0.442173 & 56.2$\pm$0.3 & $0.3\substack{ +0.1 \\ -0.08 }$ & $49.8\substack{ +7.4 \\ -5.2 }$ & $4.3\substack{ +0.9 \\ -1.0 }$ & 18.19 $\pm$ 0.14 & $11.94\substack{ +0.14 \\ -0.11 }$ & 0.45\tnote{a}\\
  PG1116$+$215 & PG1116G1 & 11:19:06.70 & $+$21:18:28.8 & 0.138114 & 138.0$\pm$0.2 & $0.76\substack{ +0.1 \\ -0.09 }$ & $26.3\substack{ +0.8 \\ -0.4 }$ & $34.4\substack{+ 0.4 \\ -0.4 }$ & 16.2 $\pm$ 0.03 & $12.0\substack{+ 0.17 \\ -0.13 }$ & 0.45\tnote{a}\\
  J113327$+$032719 & J1133G1 & 11:33:28.27 & $+$03:26:59.6 & 0.154598 & 55.6$\pm$0.1 & $0.52\substack{ +0.06 \\ -0.06 }$ & $23.5\substack{+ 0.4 \\ -0.2 }$ & $56.1\substack{ +1.7 \\ -1.3 }$ & 16.76 $\pm$ 0.96 & $11.31\substack{ +0.08 \\ -0.08 }$ & 0.20\\
  J113910$-$135043 & J1139G2 & 11:39:09.52 & $-$13:51:31.8 & 0.212259 & 174.8$\pm$0.1 & $0.98\substack{ +0.1 \\ -0.08 }$ & $85.0\substack{+ 0.1 \\ -0.6 }$ & $80.4\substack{ +0.4 \\ -0.5 }$ & 15.33 $\pm$ 0.04 & $11.96\substack{ +0.15 \\ -0.11 }$ & 0.78\\
  J113910$-$135043 & J1139G3 & 11:39:10.01 & $-$13:50:52.3 & 0.319255 & 73.3$\pm$0.4 & $0.47\substack{ +0.08 \\ -0.07 }$ & $83.4\substack{ +1.4 \\ -1.1 }$ & $39.1\substack{+1.9  \\ -1.7 }$ & 16.19 $\pm$ 0.03 & $11.74\substack{ +0.1 \\ -0.1 }$ & 0.45\tnote{a}\\
  J113910$-$135043 & J1139G4 & 11:39:11.53 & $-$13:51:08.6 & 0.204194 & 93.2$\pm$0.3 & $0.61\substack{ +0.08 \\ -0.07 }$ & $81.6\substack{+ 0.4 \\ -0.5 }$ & $5.8\substack{ +0.4 \\ -0.5 }$ & 16.28 $\pm$ 0.34 & $11.75\substack{+0.1  \\ -0.1 }$ & 0.66\\
  PG1216$+$069 & PG1216G1 & 12:19:23.44 & $+$06:38:20.1 & 0.123623 & 93.4$\pm$0.2 & $0.68\substack{ +0.07 \\ -0.07 }$ & $21.9\substack{ +18.7 \\ -21.8 }$ & $61.4\substack{ +28.5 \\ -13.4 }$ & 17.53 $\pm$ 1.47 & $11.64\substack{ +0.1 \\ -0.1 }$ & 0.41\\
  J123304$-$003134 & J1233G1 & 12:33:03.76 & $-$00:31:59.6 & 0.318757 & 88.9$\pm$0.2 & $0.55\substack{ +0.09 \\ -0.07 }$ & $38.7\substack{ +1.6 \\ -1.8 }$ & $17.0\substack{ +2.0 \\ -2.3 }$ & 15.72 $\pm$ 0.02 & $11.78\substack{ +0.11 \\ -0.1 }$ & 0.45\tnote{a}\\
  J124154$+$572107 & J1241G1 & 12:41:52.35 & $+$57:20:53.6 & 0.205267 & 21.1$\pm$0.1 & $0.16\substack{ +0.07 \\ -0.07 }$ & $56.4\substack{ +0.3 \\ -0.5 }$ & $77.6\substack{+ 0.3 \\ -0.4 }$ & 18.38 $\pm$ 0.16 & $11.55\substack{ +0.1 \\ -0.1 }$ & 0.45\tnote{a}\\
  J124154$+$572107 & J1241G2 & 12:41:52.49 & $+$57:20:42.6 & 0.217904 & 94.6$\pm$0.2 & $0.82\substack{ +0.07 \\ -0.07 }$ & $17.4\substack{ +1.4 \\ -1.6 }$ & $63.0\substack{+ 1.8 \\ -2.1 }$ & 15.59 $\pm$ 0.12 & $11.38\substack{+ 0.1 \\ -0.1 }$ & 0.29\\
  J132222$+$464546 & J1322G1 & 13:22:22.51 & $+$46:45:46.0 & 0.214431 & 38.6$\pm$0.2 & $0.16\substack{ +0.2 \\ -0.14 }$ & $57.9\substack{ +0.1 \\ -0.2 }$ & $13.9\substack{+ 0.2 \\ -0.2 }$ & 17.49 $\pm$ 0.2 & $12.32\substack{ +0.29 \\ -0.2 }$ & 0.69\\
  J134251$-$005345 & J1342G1 & 13:42:51.76 & $-$00:53:49.3 & 0.227042 & 35.3$\pm$0.2 & $0.16\substack{ +0.2 \\ -0.13 }$ & $0.1\substack{ +0.6 \\ -0 }$ & $13.2\substack{ +0.5 \\ -0.4 }$ & 18.83 $\pm$ 0.05 & $12.26\substack{+ 0.26 \\ -0.18 }$ & 0.45\tnote{a}\\
  J155504$+$362847 & J1555G1 & 15:55:05.27 & $+$36:28:48.1 & 0.189201 & 33.4$\pm$0.1 & $0.23\substack{ +0.08 \\ -0.07 }$ & $51.8\substack{ +0.7 \\ -0.7 }$ & $47.0\substack{ +0.3 \\ -0.8 }$ & 17.52 $\pm$ 0.22 & $11.69\substack{ +0.1 \\ -0.1 }$ & 0.32\\
  J213135$-$120704 & J2131G1 & 21:31:38.87 & $-$12:06:44.1 & 0.4302 & 48.4$\pm$0.2 & $0.25\substack{ +0.1 \\ -0.1 }$ & $48.3\substack{ +3.5 \\ -3.7 }$ & $14.9\substack{ +6 \\ -4.9 }$ & 19.88 $\pm$ 0.1 & $11.99\substack{ +0.16 \\ -0.12 }$ & 0.45\tnote{a}\\
  J225357$+$160853 & J2253G1 & 22:53:57.80 & $+$16:09:05.5 & 0.153718 & 31.8$\pm$0.2 & $0.25\substack{ +0.06 \\ -0.06 }$ & $33.3\substack{ +2.7 \\ -2.0 }$ & $59.6\substack{ +0.9 \\ -1.8 }$ & 16.04 $\pm$ 0.73 & $11.52\substack{ +0.1 \\ - 0.1}$ & 0.45\tnote{a}\\
  J225357$+$160853 & J2253G2 & 22:54:00.37 & $+$16:09:06.4 & 0.352787 & 203.2$\pm$0.5 & $1.61\substack{ +0.07 \\ -0.07 }$ & $36.7\substack{ +6.9 \\ -4.6 }$ & $88.7\substack{ +1.28 \\ -4.8 }$ & 14.53 $\pm$ 0.05 & $11.46\substack{+ 0.1 \\ -0.1 }$ & 0.08\\
  J225357$+$160853 & J2253G3 & 22:54:02.32 & $+$16:09:33.4 & 0.390012 & 276.3$\pm$0.2 & $1.53\substack{ +0.1 \\ -0.08 }$ & $76.1\substack{ +1.1 \\ -1.2 }$ & $24.2\substack{ +1.2 \\ -1.2 }$ & 15.19 $\pm$ 0.04 & $11.92\substack{ +0.14 \\ -0.11 }$ & 0.45\tnote{a}\\
\hline

\end{tabular}

\begin{tablenotes}
            \item[a] For galaxies without observed $g-r$ colours, we assumed an Sbc-type spectrum based on the colours found for absorbing galaxies in previous studies \citep{Steidel94, zibetti2007, Nikki13a_magiicat1, Glenn2015decmorpho}.

        \end{tablenotes}
\end{threeparttable}

\end{table}
\end{landscape}

\newpage
\section{CGM metal absorption detection}

In Table \ref{EW-detections}, we provide the rest-frame equivalent width of detected absorbers within each galaxy's CGM, and their corresponding errors. In cases where absorption is not detected for a particular galaxy, we leave that entry empty, represented by three dots (${\cdots}$). The first column presents the host galaxy ID and columns two through twelve are ordered based on the ionization potential of the ions:  {\MgII~$\lambda 2796$} (15.03~eV), {\SiII~$\lambda 1260$} (16.3~eV), 
 {\CII~$\lambda 1334$} (24.4~eV), 
 {\NII~$\lambda 1083$} (29.6~eV), 
 {\SiIII~$\lambda 1206$} (33.5~eV), {\SiIV~$\lambda 1393$} (45.1~eV), {\NIII~$\lambda 989$} (47.5~eV), {\CIII~$\lambda 977$} (47.9~eV), {\CIV~$\lambda 1548$} (64.5~eV), {\NV~$\lambda 1238$} (97.9~eV), and {\OVI~$\lambda 1031$} (138.1~eV).

\begin{table*}
\centering

\renewcommand{\arraystretch}{1.5}
\setlength{\tabcolsep}{4pt}
\caption{Absorption EW and detection}
\label{EW-detections}

\begin{tabular}{l c c c c c c c c c c c}
\hline
\hline

\multicolumn{1}{l}{Galaxy} &
  \multicolumn{1}{c}{$W_{r}({\MgII})$} &
  \multicolumn{1}{c}{$W_{r}({\SiII})$} &
  \multicolumn{1}{c}{$W_{r}({\CII})$} &
  \multicolumn{1}{c}{$W_{r}({\NII})$} &
  \multicolumn{1}{c}{$W_{r}({\SiIII})$} &
  \multicolumn{1}{c}{$W_{r}({\SiIV})$} &
  \multicolumn{1}{c}{$W_{r}({\NIII})$} &
  \multicolumn{1}{c}{$W_{r}({\CIII})$} &
  \multicolumn{1}{c}{$W_{r}({\CIV})$} &
  \multicolumn{1}{c}{$W_{r}({\NV})$} &
  \multicolumn{1}{c}{$W_{r}({\OVI})$} \\
\hline

J0351G1 & 0.20 $\pm$ 0.04 & 0.13 $\pm$ 0.04 & 0.19 $\pm$ 0.04 & ... & 0.44 $\pm$ 0.03 & 0.17 $\pm$ 0.04 & ... & ... & 0.62 $\pm$ 0.11 & 0.15 $\pm$ 0.03 & 0.38 $\pm$ 0.02\\
  J0407G1 & ... & ... & ... & ... & ... & ... & ... & 0.09 $\pm$ 0.0 & ... & ... & 0.22 $\pm$ 0.01\\
  J0456G1 & ... & ... & 0.02 $\pm$ 0.01 & ... & 0.11 $\pm$ 0.03 & 0.06 $\pm$ 0.0 & ... & ... & 0.21 $\pm$ 0.06 & 0.04 $\pm$ 0.03 & 0.21 $\pm$ 0.01\\
  J0456G2 & ... & ... & ... & ... & 0.02 $\pm$ 0.02 & ... & ... & ... & 0.28 $\pm$ 0.04 & 0.03 $\pm$ 0.02 & ...\\
  J0853G2 & ... & 0.29 $\pm$ 0.02 & 0.31 $\pm$ 0.01 & 0.14 $\pm$ 0.01 & ... & 0.05 $\pm$ 0.07 & 0.09 $\pm$ 0.02 & 0.55 $\pm$ 0.02 & ... & ... & ...\\
  J0914G1 & ... & 0.03 $\pm$ 0.05 & 0.04 $\pm$ 0.04 & ... & 0.09 $\pm$ 0.03 & ... & 0.07 $\pm$ 0.03 & ... & ... & ... & 0.34 $\pm$ 0.04\\
  J0943G1 & ... & ... & ... & ... & ... & ... & ... & 0.5 $\pm$ 0.05 & ... & 0.12 $\pm$ 0.12 & 0.24 $\pm$ 0.06\\
  J0950G1 & 0.65 $\pm$ 0.04 & 0.36 $\pm$ 0.05 & 0.41 $\pm$ 0.08 & 0.23 $\pm$ 0.02 & 0.52 $\pm$ 0.04 & 0.2 $\pm$ 0.09 & 0.24 $\pm$ 0.02 & 0.64 $\pm$ 0.02 & ... & ... & 0.11 $\pm$ 0.02\\
  PG1001G1 & ... & ... & ... & ... & ... & ... & ... & ... & ... & ... & 0.11 $\pm$ 0.01\\
  J1009G1 & 0.09 $\pm$ 0.01 & 0.08 $\pm$ 0.01 & 0.15 $\pm$ 0.01 & ... & 0.26 $\pm$ 0.01 & 0.14 $\pm$ 0.02 & ... & ... & ... & 0.16 $\pm$ 0.01 & 0.55 $\pm$ 0.04\\
  J1041G1 & 0.68 $\pm$ 0.04 & 0.37 $\pm$ 0.04 & 0.48 $\pm$ 0.04 & 0.18 $\pm$ 0.03 & 0.53 $\pm$ 0.04 & ... & 0.32 $\pm$ 0.02 & 0.57 $\pm$ 0.01 & ... & ... & 0.32 $\pm$ 0.03\\
  PG1116G1 & 0.09 $\pm$ 0.01 & 0.05 $\pm$ 0.0 & 0.08 $\pm$ 0.01 & 0.03 $\pm$ 0.01 & 0.09 $\pm$ 0.01 & 0.02 $\pm$ 0.03 & 0.04 $\pm$ 0.0 & 0.11 $\pm$ 0.0 & 0.02 $\pm$ 0.02 & ... & 0.07 $\pm$ 0.01\\
  J1133G1 & ... & 0.07 $\pm$ 0.05 & 0.09 $\pm$ 0.04 & ... & 0.1 $\pm$ 0.04 & 0.04 $\pm$ 0.1 & ... & ... & ... & ... & 0.21 $\pm$ 0.04\\
  J1139G2 & ... & ... & ... & ... & 0.03 $\pm$ 0.03 & ... & ... & 0.07 $\pm$ 0.01 & ... & ... & 0.04 $\pm$ 0.01\\
  J1139G3 & ... & ... & 0.04 $\pm$ 0.03 & ... & ... & ... & ... & 0.31 $\pm$ 0.02 & ... & ... & 0.22 $\pm$ 0.02\\
  J1139G4 & ... & ... & 0.11 $\pm$ 0.04 & ... & 0.43 $\pm$ 0.05 & ... & 0.09 $\pm$ 0.02 & 0.52 $\pm$ 0.02 & ... & ... & 0.24 $\pm$ 0.01\\
  PG1216G1 & ... & ... & ... & ... & 0.09 $\pm$ 0.01 & ... & 0.12 $\pm$ 0.04 & 0.59 $\pm$ 0.04 & 0.5 $\pm$ 0.02 & 0.09 $\pm$ 0.01 & 0.38 $\pm$ 0.08\\
  J1233G1 & ... & ... & 0.05 $\pm$ 0.08 & ... & 0.12 $\pm$ 0.05 & ... & ... & 0.47 $\pm$ 0.02 & ... & ... & 0.4 $\pm$ 0.03\\
  J1241G1 & 1.0 $\pm$ 0.04 & 0.54 $\pm$ 0.04 & 0.63 $\pm$ 0.04 & ... & 0.62 $\pm$ 0.03 & 0.52 $\pm$ 0.06 & 0.43 $\pm$ 0.02 & 0.65 $\pm$ 0.02 & ... & 0.1 $\pm$ 0.03 & 0.5 $\pm$ 0.02\\
  J1241G2 & ... & ... & ... & ... & 0.07 $\pm$ 0.02 & ... & ... & 0.25 $\pm$ 0.01 & ... & 0.08 $\pm$ 0.03 & 0.36 $\pm$ 0.02\\
  J1322G1 & 0.3 $\pm$ 0.04 & 0.23 $\pm$ 0.06 & 0.31 $\pm$ 0.1 & ... & 0.58 $\pm$ 0.04 & 0.11 $\pm$ 0.07 & 0.17 $\pm$ 0.03 & 0.6 $\pm$ 0.02 & ... & ... & 0.35 $\pm$ 0.02\\
  J1342G1 & 1.4 $\pm$ 0.04 & 0.71 $\pm$ 0.05 & 0.86 $\pm$ 0.05 & 0.46 $\pm$ 0.03 & 0.89 $\pm$ 0.04 & 0.24 $\pm$ 0.09 & 0.44 $\pm$ 6.6 & 1.1 $\pm$ 0.03 & ... & ... & 0.29 $\pm$ 0.03\\
  J1555G1 & 0.31 $\pm$ 0.04 & 0.2 $\pm$ 0.05 & 0.31 $\pm$ 0.09 & 0.1 $\pm$ 0.05 & 0.43 $\pm$ 0.05 & 0.22 $\pm$ 0.07 & 0.24 $\pm$ 0.04 & 0.54 $\pm$ 0.03 & ... & ... & 0.31 $\pm$ 0.03\\
  J2131G1 & 0.45 $\pm$ 0.04 & 0.26 $\pm$ 0.01 & 0.33 $\pm$ 0.02 & 0.16 $\pm$ 0.01 & 0.41 $\pm$ 0.03 & ... & ... & 0.5 $\pm$ 0.01 & 0.33 $\pm$ 0.09 & ... & 0.31 $\pm$ 0.01\\
  J2253G1 & ... & 0.06 $\pm$ 0.02 & 0.14 $\pm$ 0.02 & ... & 0.29 $\pm$ 0.05 & 0.21 $\pm$ 0.16 & ... & ... & 0.36 $\pm$ 0.06 & ... & 0.18 $\pm$ 0.05\\
  J2253G2 & ... & ... & ... & ... & ... & ... & ... & 0.34 $\pm$ 0.03 & ... & 0.07 $\pm$ 0.04 & 0.31 $\pm$ 0.05\\
  J2253G3 & ... & 0.02 $\pm$ 0.04 & ... & ... & 0.06 $\pm$ 0.04 & ... & 0.04 $\pm$ 0.04 & 0.27 $\pm$ 0.03 & ... & 0.04 $\pm$ 0.03 & 0.15 $\pm$ 0.04\\
\hline
\end{tabular}
\end{table*}

\newpage
\section{EW co-rotation fraction measurements}

In Table~\ref{EW corot}, we provide the EW co-rotation fraction (\fewcorot) for all the detected ions in each system, where typical uncertainties on {\fewcorot} have a range of $0.001-0006$. The EW co-rotation fraction of {\HI} is adopted from \citetalias{GF1}. The last column presents the slope of the {\fewcorot} vs. ionization with $1\sigma$ uncertainties (\slope).  Slopes are not reported for galaxies with fewer than three metal lines or where there is very little gas consistent with co-rotation.

\begin{table*}

\centering

\renewcommand{\arraystretch}{1.4}
\setlength{\tabcolsep}{4pt}
\caption{Absorption EW co-rotation fraction}
\label{EW corot}

\begin{threeparttable}
\begin{tabular}{l c c c c c c c c c c c c c}
\hline
\hline

& \multicolumn{12}{c}{EW co-rotation fraction ({\fewcorot})\tnote{a}} \\
\cmidrule(lr){2-13}

  \multicolumn{1}{c}{Galaxy} &
  \multicolumn{1}{c}{\HI} &
  \multicolumn{1}{c}{\MgII} &
  \multicolumn{1}{c}{\SiII} &
  \multicolumn{1}{c}{\CII} &
  \multicolumn{1}{c}{\NII} &
  \multicolumn{1}{c}{\SiIII} &
  \multicolumn{1}{c}{\SiIV} &
  \multicolumn{1}{c}{\NIII} &
  \multicolumn{1}{c}{\CIII} &
  \multicolumn{1}{c}{\CIV} &
  \multicolumn{1}{c}{\NV} &
  \multicolumn{1}{c}{\OVI} &
  \multicolumn{1}{c}{\slope [$\times 10^{-4}$]} \\

\hline
  J0351G1 & 0.69 & 0.32 & 0.45 & 0.45 & ... & 0.65 & 0.45 & ... & ... & 0.68 & 0.76 & 0.73 & 29.8 $\pm$ 8.9\\
  J0407G1 & 0.57 & ... & ... & ... & ... & ... & ... & ... & 0.72 & ... & ... & 0.6 & ...\tnote{b}   \\
  J0456G1 & 0.78 & ... & ... & 0.8 & ... & 0.87 & 0.84 & ... & ... & 0.91 & 0.65 & 0.51 & $-$30.1 $\pm$ 8.8\\
  J0456G2 & 0.59 & ... & ... & ... & ... & 0.62 & ... & ... & ... & 0.63 & 0.66 & ... & 6.4 $\pm$ 2.4\\
  J0853G2 & 0.57 & ... & 0.97 & 0.95 & 0.97 & ... & 0.97 & 0.97 & 0.9 & ... & ... & ... & $-$6.5 $\pm$ 9.9\\
  J0914G1 & 0.23 & ... & 0.23 & 0.14 & ... & 0.19 & ... & 0.07 & ... & ... & ... & 0.16 & $-$1.8 $\pm$ 7.1\\
  J0943G1 & 0.93 & ... & ... & ... & ... & ... & ... & ... & 0.99 & ... & 0.97 & 0.97 & $-$3.0 $\pm$ 1.0\\
  J0950G1 & 0.29 & 0.04 & 0.1 & 0.09 & 0.05 & 0.16 & 0.06 & 0.07 & 0.21 & ... & ... & 0.1 & 2.4 $\pm$ 5.4\\
  PG1001G1 & 0.64 & ... & ... & ... & ... & ... & ... & ... & ... & ... & ... & 0.98 & ...\tnote{b}  \\
  J1009G1 & 0.52 & 0.02 & 0.1 & 0.36 & ... & 0.31 & 0.13 & ... & ... & ... & 0.49 & 0.56 & 36.2 $\pm$ 10.9\\
  J1041G1 & 0.94 & 1.0 & 1.0 & 1.0 & 1.0 & 1.0 & ... & 1.0 & 0.99 & ... & ... & 0.99 & $-$0.9 $\pm$ 0.2\\
  PG1116G1 & 0.59 & 1.0 & 1.0 & 1.0 & 1.0 & 1.0 & 1.0 & 1.0 & 1.0 & 1.0 & ... & 1.0 & $-$0.2  $\pm$ 0.1\\
  J1133G1 & 1.0 & ... & 0.99 & 1.0 & ... & 0.99 & 0.99 & ... & ... & ... & ... & 1.0 & 0.2 $\pm$ 0.4\\
  J1139G2 & 0.03 & ... & ... & ... & ... & 0.01 & ... & ... & 0.01 & ... & ... & 0.02 & ...\tnote{c}   \\
  J1139G3 & 0.46 & ... & ... & 0.81 & ... & ... & ... & ... & 0.55 & ... & ... & 0.6 & $-$11.4 $\pm$ 19.1\\
  J1139G4 & 0.68 & ... & ... & 0.51 & ... & 0.63 & ... & 0.63 & 0.68 & ... & ... & 0.68 & 9.4 $\pm$ 7.0\\
  PG1216G1 & 0.17 & ... & ... & ... & ... & 0.34 & ... & 0.31 & 0.23 & 0.19 & 0.3 & 0.21 & $-$7.0 $\pm$ 7.0\\
  J1233G1 & 0.43 & ... & ... & 0.31 & ... & 0.44 & ... & ... & 0.31 & ... & ... & 0.24 & $-$10.6 $\pm$ 8.0\\
  J1241G1 & 0.81 & 1.0 & 0.98 & 0.95 & ... & 0.94 & 0.97 & 0.96 & 0.79 & ... & 0.93 & 0.86 & $-$8.4 $\pm$ 5.3\\
  J1241G2 & 0.27 & ... & ... & ... & ... & 0.35 & ... & ... & 0.15 & ... & 0.21 & 0.24 & $-$4.2 $\pm$ 11.6\\
  J1322G1 & 0.58 & 0.52 & 0.48 & 0.54 & ... & 0.5 & 0.37 & 0.49 & 0.48 & ... & ... & 0.37 & $-$11.4 $\pm$ 4.6\\
  J1342G1 & 0.39 & 0.21 & 0.21 & 0.27 & 0.24 & 0.28 & 0.25 & 0.26 & 0.27 & ... & ... & 0.3 & 5.6 $\pm$ 2.2\\
  J1555G1 & 0.64 & 0.92 & 0.83 & 0.85 & 0.83 & 0.87 & 0.91 & 0.91 & 0.82 & ... & ... & 0.76 & $-$9.1 $\pm$ 4.0\\
  J2131G1 & 0.58 & 0.97 & 0.95 & 0.93 & 0.94 & 0.83 & ... & ... & 0.76 & 0.94 & ... & 0.66 & $-$22.4 $\pm$ 6.6\\
  J2253G1 & 0.45 & ... & 0.31 & 0.42 & ... & 0.3 & 0.27 & ... & ... & 0.33 & ... & 0.29 & $-$4.4 $\pm$ 5.7\\
  J2253G2 & 0.74 & ... & ... & ... & ... & ... & ... & ... & 0.96 & ... & 0.8 & 0.67 & $-$31.4 $\pm$ 0.4\\
  J2253G3 & 0.01 & ... & 0.0 & ... & ... & 0.0 & ... & 0.0 & 0.0 & ... & 0.0 & 0.0 & ...\tnote{c}  \\

\hline\end{tabular}

\begin{tablenotes}
            \item[a] Typical uncertainties are on the order of $0.001-0.006$ as determined from a bootstrap analysis where we varied the galaxy and absorption redshifts within their error bars.
            \item[b] Fewer than three metal lines were detected. A slope cannot be reliably measured.
            \item[c] No significant co-rotation was detected for this system.
        \end{tablenotes}
\end{threeparttable}

\end{table*}






\bsp	
\label{lastpage}
\end{document}